\documentclass[sigconf]{acmart}
\usepackage{microtype}
\usepackage[american]{babel}
\usepackage{amsfonts}
\usepackage{amsmath}
\usepackage{multirow}
\usepackage{booktabs} 
\usepackage{bm}

\AtBeginDocument{%
  \providecommand\BibTeX{{%
    \normalfont B\kern-0.5em{\scshape i\kern-0.25em b}\kern-0.8em\TeX}}}

\copyrightyear{2024} 
\acmYear{2024} 
\setcopyright{acmlicensed}
\acmConference[KDD '24]{Proceedings of the 30th ACM SIGKDD Conference on Knowledge Discovery and Data Mining}{August 25--29, 2024}{Barcelona, Spain}
\acmBooktitle{Proceedings of the 30th ACM SIGKDD Conference on Knowledge Discovery and Data Mining (KDD '24), August 25--29, 2024, Barcelona, Spain}
\acmDOI{10.1145/3637528.3671988}
\acmISBN{979-8-4007-0490-1/24/08}
\begin{document}

\title{ORCDF: An Oversmoothing-Resistant Cognitive Diagnosis Framework for Student Learning in Online Education Systems}

\author{Hong Qian}
\email{hqian@cs.ecnu.edu.cn}
\orcid{0000-0003-2170-5264}
\affiliation{%
  \institution{School of Computer Science and Technology, and Shanghai Institute of AI Education\\ East China Normal University}
  \city{Shanghai}
  \country{China}
}

\author{Shuo Liu}
\email{shuoliu@stu.ecnu.edu.cn}
\orcid{0000-0001-7970-3187}
\affiliation{%
  \institution{School of Computer Science and Technology\\ East China Normal University}
  \city{Shanghai}
  \country{China}
}

\author{Mingjia Li}
\email{52275901007@stu.ecnu.edu.cn}
\orcid{0000-0001-5740-0302}
\affiliation{%
  \institution{School of Computer Science and Technology, and Shanghai Institute of AI Education\\ East China Normal University}
  \city{Shanghai}
  \country{China}
}

\author{Bingdong Li}
\email{bdli@cs.ecnu.edu.cn}
\orcid{0000-0001-5740-0302}
\affiliation{%
  \institution{School of Computer Science and Technology, and Shanghai Institute of AI Education\\ East China Normal University}
  \city{Shanghai}
  \country{China}
}

\author{Zhi Liu}
\email{zhliu@cs.ecnu.edu.cn}
\orcid{0000-0001-5740-0302}
\affiliation{%
  \institution{School of Computer Science and Technology, and Shanghai Institute of AI Education\\ East China Normal University}
  \city{Shanghai}
  \country{China}
}

\author{Aimin Zhou}
\authornote{Aimin Zhou is the corresponding author.}
\email{amzhou@cs.ecnu.edu.cn}
\orcid{0000-0002-4768-5946}
\affiliation{%
  \institution{School of Computer Science and Technology, and Shanghai Institute of AI Education\\ East China Normal University}
  \city{Shanghai}
  \country{China}
}



\begin{abstract}
Cognitive diagnosis models (CDMs) are designed to learn students' mastery levels using their response logs. CDMs play a fundamental role in online education systems since they significantly influence downstream applications such as teachers' guidance and computerized adaptive testing. Despite the success achieved by existing CDMs, we find that they suffer from a thorny issue that the learned students' mastery levels are too similar. This issue, which we refer to as oversmoothing, could diminish the CDMs' effectiveness in downstream tasks. CDMs comprise two core parts: learning students' mastery levels and assessing mastery levels by fitting the response logs. This paper contends that the oversmoothing issue arises from that existing CDMs seldom utilize response signals on exercises in the learning part but only use them as labels in the assessing part. To this end, this paper proposes an oversmoothing-resistant cognitive diagnosis framework (ORCDF) to enhance existing CDMs by utilizing response signals in the learning part. Specifically, ORCDF introduces a novel response graph to inherently incorporate response signals as types of edges. Then, ORCDF designs a tailored response-aware graph convolution network (RGC) that effectively captures the crucial response signals within the response graph. Via ORCDF, existing CDMs are enhanced by replacing the input embeddings with the outcome of RGC, allowing for the consideration of response signals on exercises in the learning part. Extensive experiments on real-world datasets show that ORCDF not only helps existing CDMs alleviate the oversmoothing issue but also significantly enhances the models' prediction and interpretability performance. Moreover, the effectiveness of ORCDF is validated in the downstream task of computerized adaptive testing.
\end{abstract}




\maketitle

\section{Introduction}
\begin{figure}[!t]
\centering
\includegraphics[width=0.90\linewidth]{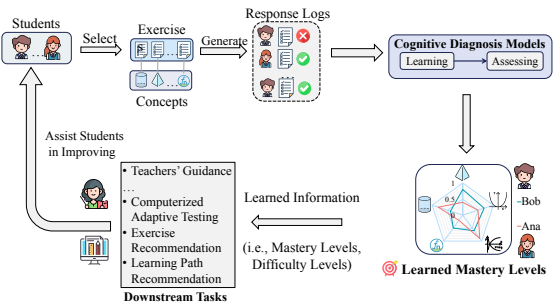}
\caption{An example of CD, as well as relationships between CD and downstream tasks.}
\label{fig:CDA}
\end{figure}

Cognitive diagnosis (CD)~\cite{Liu2023New} serves as the foundational element in online intelligent education systems. It exerts an upstream and fundamental influence on subsequent modules such as computer adaptive testing~\cite{ncat}, course recommendation~\cite{xu2020course, Jiang2023Rk} and learning path suggestions~\cite{Sun2023AKT, Shen2024Survey}, among others. Specifically, as illustrated in Figure~\ref{fig:CDA}, CD aims to learn students' underlying mastery levels (Mas) by analyzing their historical response logs, thereby providing insights into various attributes of exercises, such as difficulty level (Diff) and discrimination (Disc). In recent years, an array of cognitive diagnosis models (CDMs) have emerged, prominently featuring frameworks such as item response theory (IRT)~\cite{IRTA} and the neural cognitive diagnosis model (NCDM)~\cite{NCDM}. The two core parts of CDM include learning students' Mas and assessing the learned Mas by fitting the response logs. The function used in the latter part is often referred to as the interaction function (IF). IRT utilizes a latent factor to represent Mas and adopts the logistic function as IF. In contrast, NCDM replaces the traditional IFs with multi-layer perceptrons (MLP) and uses concept-specific vectors (i.e., set the embedding dimension being equal to the number of concepts) to characterize Mas. As embedding-based methods rapidly evolve and gain prominence, there is an increasing trend of representing both students and exercises in a vectorized form, and they are gradually refined by using a variety of advanced techniques~\cite{kscd,hiercdf,RCD,KANCD,Li2022CDMFKC, Shao2024ASGCD}.

\begin{figure}[!t]
\centering
\includegraphics[width=0.55\linewidth]{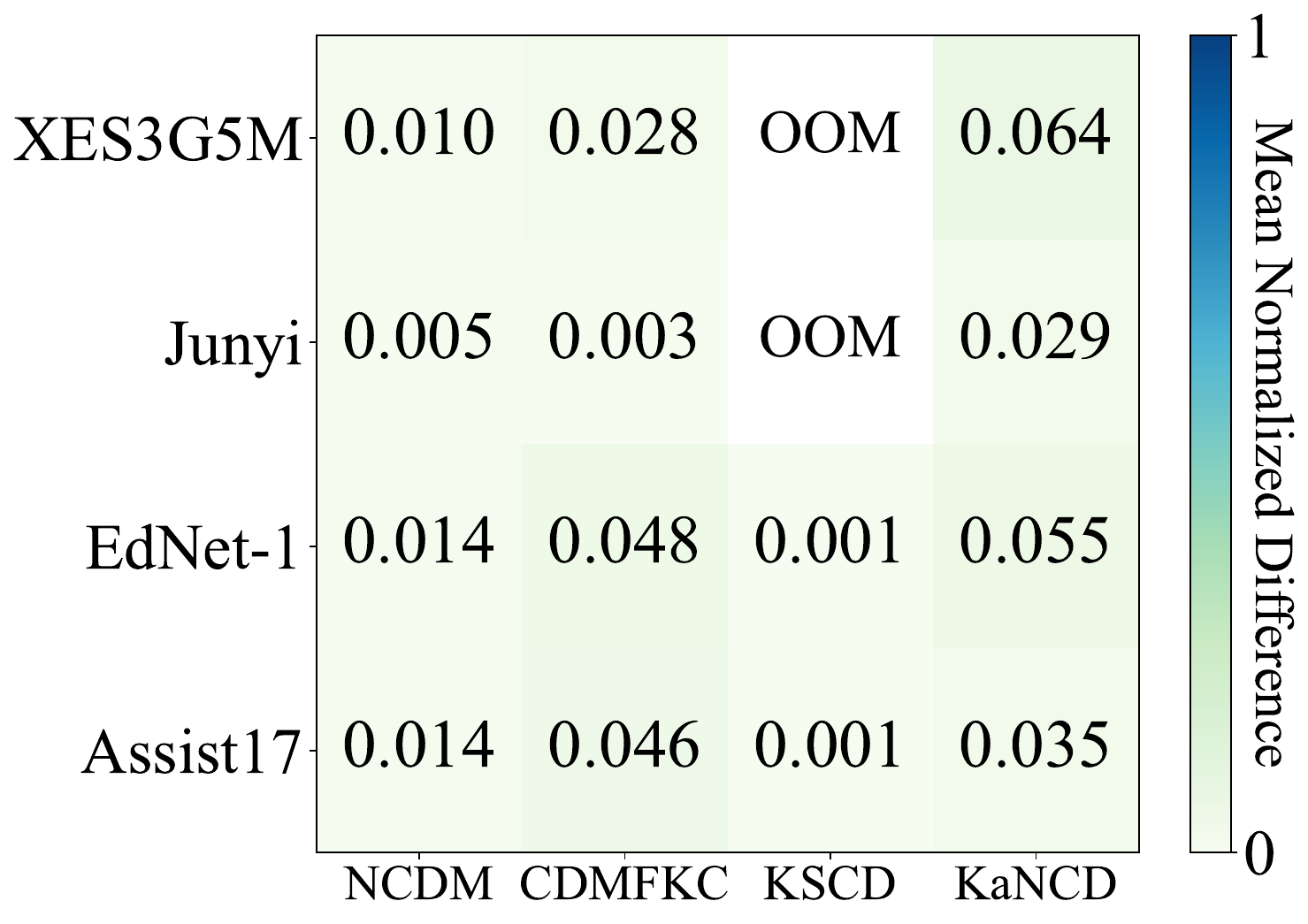}
\caption{Results of motivation and pilot experiments: the oversmoothing issue in most existing CDMs is highlighted. The degree of oversmoothing is measured by the mean normalized difference (the lower the worse). OOM means the out-of-memory on an NVIDIA 3090 GPU. The vertical axis represents the names of four real-world datasets, and the horizontal axis lists the representative existing CDMs.}
\label{fig:oversmoothing}
\end{figure}

Despite the success, \textit{\textbf{this paper, for the first time, identifies that existing CDMs share a potential and thorny issue that the learned Mas of students are too similar}}. We refer to this issue as oversmoothing. Oversmoothing could diminish the CDMs' effectiveness in down-stream tasks. To support the motivation of this paper and reveal the oversmoothing issue, we conduct a pilot study on four real-world datasets collected from the online education systems, ensuring a diverse range of circumstances in the students' response logs. To characterize the degree of oversmoothing, inspired by~\cite{Li2018Oversmoothing}, the mean normalized difference (MND) is proposed to measure the Mas learned by CDMs. Intuitively, the larger the MND value, the bigger the difference among students' Mas that learned by CDMs. Details of MND are elaborated in Section~\ref{eq:oversmoothing}. As shown in Figure~\ref{fig:oversmoothing}, although CDMs such as NCDM~\cite{NCDM}, CDMFKC~\cite{Li2022CDMFKC}, KSCD~\cite{kscd} and KaNCD~\cite{KANCD} achieve commendable prediction performance, the MND values of Mas they have learned are quite small and hard to distinguish. Since CD is an upstream task, addressing this issue is urgent. For instance, if teachers rely on the outcomes of CD to assist student development, exceedingly subtle distinctions could lead to confusion. Intuitively, if MND is 0.005, it implies that the average difference in Mas for two students in a class on certain concepts is merely 0.005 (e.g., 0.51 and 0.515). Such a small margin could potentially bring difficulty to teachers to accurately assess the cognitive state of entire class. This not only fails to aid students but could also result in misguided instruction. Moreover, for downstream algorithms, a diagnosis result
plagued by oversmoothing may lead to erroneous recommendations of learning materials, causing irreversible impacts on students.

One straightforward approach is to design a regularization term aimed at amplifying the differences between students. However, achieving a balance between the weight of this regularization term and the binary cross-entropy (BCE) loss during training is challenging. Besides, although this direct approach may help in mitigating the oversmoothing issue, it could compromise the model's prediction performance, since it forcefully amplifies the differences among all students and adversely affects the learning of students' Mas who should, in principle, be closely aligned. In this paper, we contend that the oversmoothing issue arises because existing CDMs seldom utilize response signals in the learning part but only use them as labels in the assessing part. For instance, students with right response on exercises with high difficulty levels should attain higher Mas on corresponding concepts in the learning part. Cooperating response signals in both learning and assessing parts of CDMs can widen the gap among students' Mas as they reserve the unique feature in students' response logs. 

To this end, this paper proposes an oversmoothing-resistant cognitive diagnosis framework (ORCDF) to enhance existing CDMs by utilizing response signals in the learning part. Specifically, ORCDF introduces a novel response graph, which utilizes response logs and a $\mathbf{Q}$-matrix, inherently incorporating response signals as types of edges. Then, ORCDF designs a tailored response-aware graph convolution network (RGC) that effectively captures the crucial response signals within the response graph. We reveal that by utilizing the multiple layers of RGC, we achieve a multi-perspective analysis of student mastery. This is accomplished by combining the outcomes from multiple layers of RGC, leading to a more comprehensive understanding of student learning. Via ORCDF, existing CDMs are enhanced by replacing the input embeddings with the outcome of RGC through the transformation layer, allowing for the consideration of response signals on exercises in the learning part.
Nevertheless, ORCDF encounters a new challenge: overemphasizing the role of response signals can exacerbate the guess and slip problem. This problem occurs when students guess in order to answer correctly or make mistakes on exercises they actually master, and could potentially lead models to make unreasonable inference of students' Mas. Different from previous methods that introduce extra parameters for guess and slip probabilities~\cite{fuzzycdm}, this paper addresses the guess and slip problem in student-exercise interactions by considering them as noise edges in the response graph. Specifically, we flip the student-exercise edge in the response graph with a flip ratio $p_f$ (i.e., changing right to wrong, and wrong to right) in each epoch during the learning phase. We then design a loss function that ensures consistency in learning despite the presence of different noises, thereby mitigating the guess and slip problem. Extensive experiments
show ORCDF's superiority over state-of-the-art methods in terms of resisting oversmoothing, enhancing prediction performance, and improving interpretability. Finally, we validate the efficacy of ORCDF in downstream tasks.




The subsequent sections respectively recap the related work, present the preliminaries, introduce the proposed ORCDF, show the
empirical analysis and finally conclude the paper.

\section{Related Work}
\textbf{Cognitive Diagnosis Models.} CDMs involve various approaches, such as latent factor models like IRT and MIRT (multidimensional IRT), or concept mastery pattern models like the deterministic input, noisy and gate (DINA) model, to infer students' mastery levels. DINA, a classic CDM, employs binary variables to represent mastery levels where 0 means unmastered and 1 means mastered. However, recent advances in deep learning have led to significant improvements in handling large-scale interactions. Notably, NCDM uses MLP as its IF, treating mastery patterns as continuous variables ranging from 0 to 1. This evolution in approach has been paralleled by diverse methods in analyzing response logs, including MLP based~\cite{Li2022CDMFKC, kscd, FineCD}, graph attention networks~\cite{icd} and Bayesian networks~\cite{hiercdf, Wu2023Bns}, each contributing to a more nuanced understanding of student learning patterns. However, as depicted in Figure~\ref{fig:oversmoothing}, these advanced CDMs encounter the oversmoothing issue which could potentially hinder the application of CD in downstream tasks of intelligent education, affecting their performance and consequently
impacting student learning. To the best of our knowledge, the oversmoothing issue in the field of CD remains unexplored.

\textbf{Oversmoothing Issue.} The oversmoothing issue~\cite{Li2018Oversmoothing} is a significant problem in graph representation learning (GRL). Many studies  have shown that the layers of graph neural network (GNN) deepen, the representations of graph nodes become increasingly smooth, leading to a substantial decrease in accuracy. This has prompted numerous researchers to employ a variety of methods~\cite{Min2022ScatterGCN} to address this issue, enabling deeper GNN architectures. The same phenomenon is also observed in various fields where graphs are used for data mining. For example, in recommendation systems, graph collaborative filtering (GCF)~\cite{Xia2022HCCF} faces the oversmoothing problem, which arises for the same reasons as in GRL due to the stacking of GNN layers. However, in the context of CD, oversmoothing is not a result of
stacking GNN layers, since most CDMs like NCDM, CDMFKC, KSCD and KaNCD do not utilize GNN. Yet, this issue does exist and is urgent, as shown in Figure~\ref{fig:oversmoothing}. Thus, existing solutions to addressing oversmoothing in GRL and GCF are not suitable to resolve the oversmoothing issue in CD.

\section{Preliminaries}
This section first introduces the fundamental elements of CD and then introduces the formal problem definition of CD and oversmoothing issue in CDMs. We also give abbreviations for terms in Table~\ref{tab:abbreviations} at the beginning of the Appendix.

\textbf{Cognitive Diagnosis Basis.} Consider an education scenario which contains three sets: $S = \{s_1, \ldots, s_N\}$, $E = \{e_1, \ldots, e_M\}$, and $C = \{c_1, \ldots, c_Z\}$. They symbolize students, exercises and knowledge concepts, with respective sizes of $N$, $M$ and $Z$. $Q$ represents the relationship between exercises and knowledge concepts, which can be regarded as a binary matrix $\mathbf{Q} = (\mathbf{Q}_{iz})_{M \times Z}$, where $\mathbf{Q}_{iz} \in \{0, 1\}$ means whether $e_i$ relates to $c_z$ or not. Students from set $S$, driven by unique interests and requirements, select exercises from $E$. The results are documented as response logs. Specifically, these logs can be illustrated as triplets $T = \{(s, e, r)\ | \ s \in S, e \in E, r_{se} \in \{0, 1\}\}$. $r_{se} = 1$ represents correct and $r_{se} = 0$ represents wrong. In this paper, we treat response logs as interaction matrix $\mathbf{I} \in \mathbb{R}^{N \times M}$. It contains three categorical values ($1$ means right, $0$ means no interaction and $-1$ means wrong). Finally, we give the formal definition of the CD task and oversmoothing issue in CDMs.

\begin{definition}[Problem Definition]
\textit{Given interaction matrix $\mathbf{I} \in \mathbb{R}^{N \times M}$, a binary matrix $\mathbf{Q} \in \mathbb{R}^{M \times Z}$, the goal of cognitive diagnosis is to infer $\mathbf{Mas} \in \mathbb{R}^{N \times Z}$, which denotes the latent mastery level of students on each concept}.
\end{definition}

\begin{definition}[Oversmoothing in CDMs]
\textit{Given the learned  $\mathbf{Mas} \in \mathbb{R}^{N \times Z}$ by CDMs, if the difference in students' $\mathbf{Mas}$ is sufficiently small, it indicates the presence of oversmoothing issue in CDMs.}
\end{definition}
In this paper, we utilize the mean normalized difference proposed in Section~\ref{eq:oversmoothing} to quantify the degree of oversmoothing.

\begin{figure*}[!t]
\centering
\includegraphics[width=0.99\linewidth]{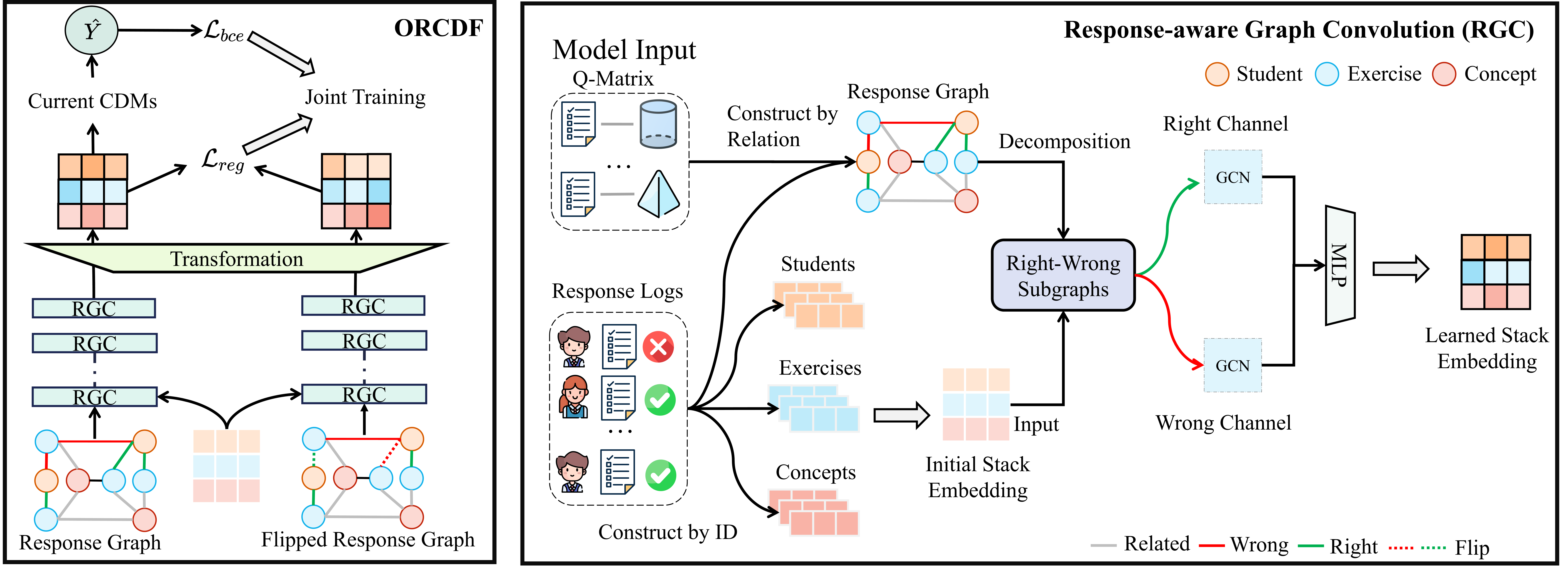}
\caption{The left side provides an overview of the proposed ORCDF. The right side details the main component of ORCDF.}
\label{fig:framework}
\end{figure*}

\begin{figure}[!h]
\centering
\begin{minipage}{0.495\linewidth}\centering
    \includegraphics[width=1.0\textwidth]{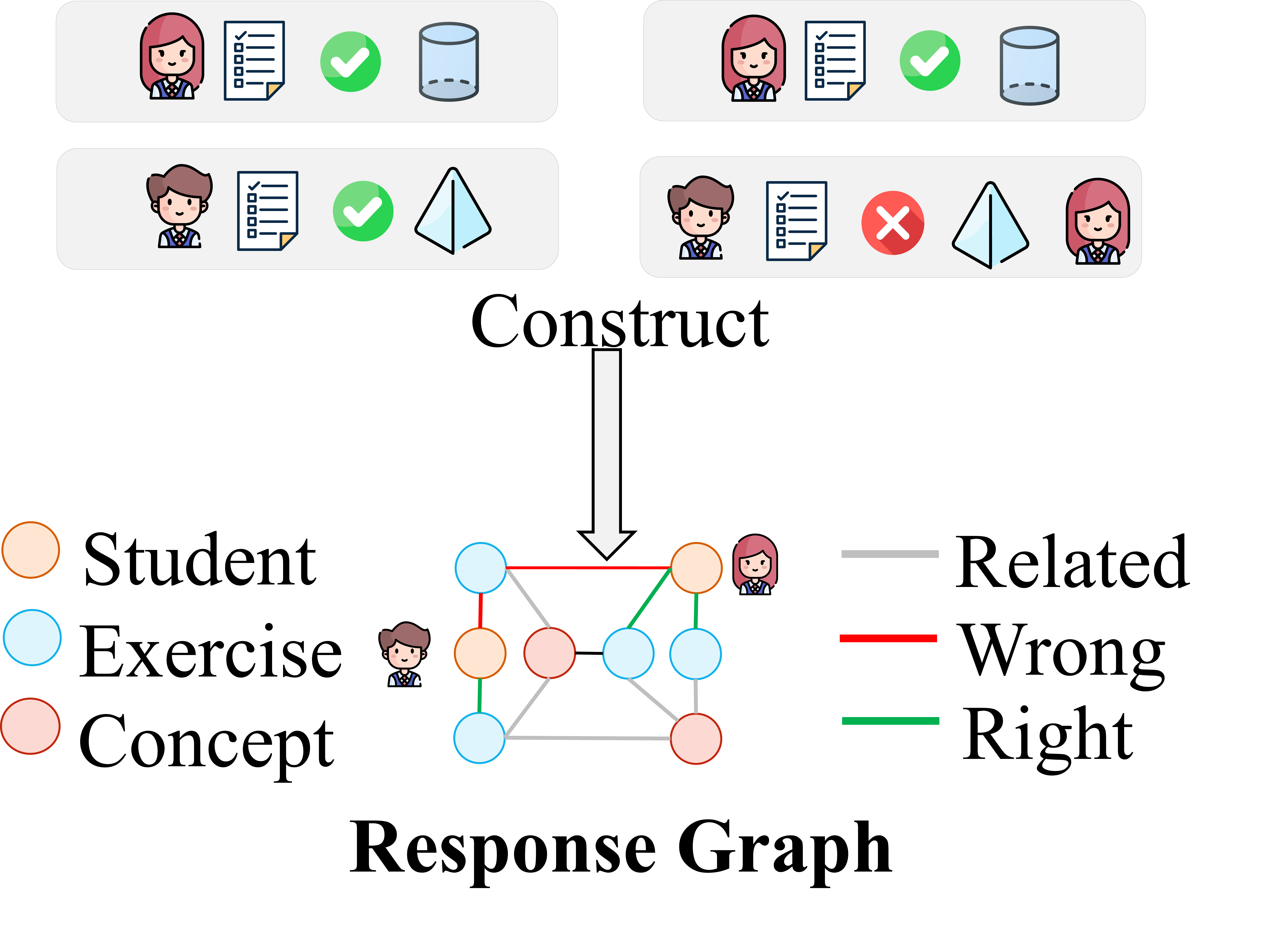}\\
    \end{minipage}
\begin{minipage}{0.495\linewidth}\centering
    \includegraphics[width=0.7\textwidth]{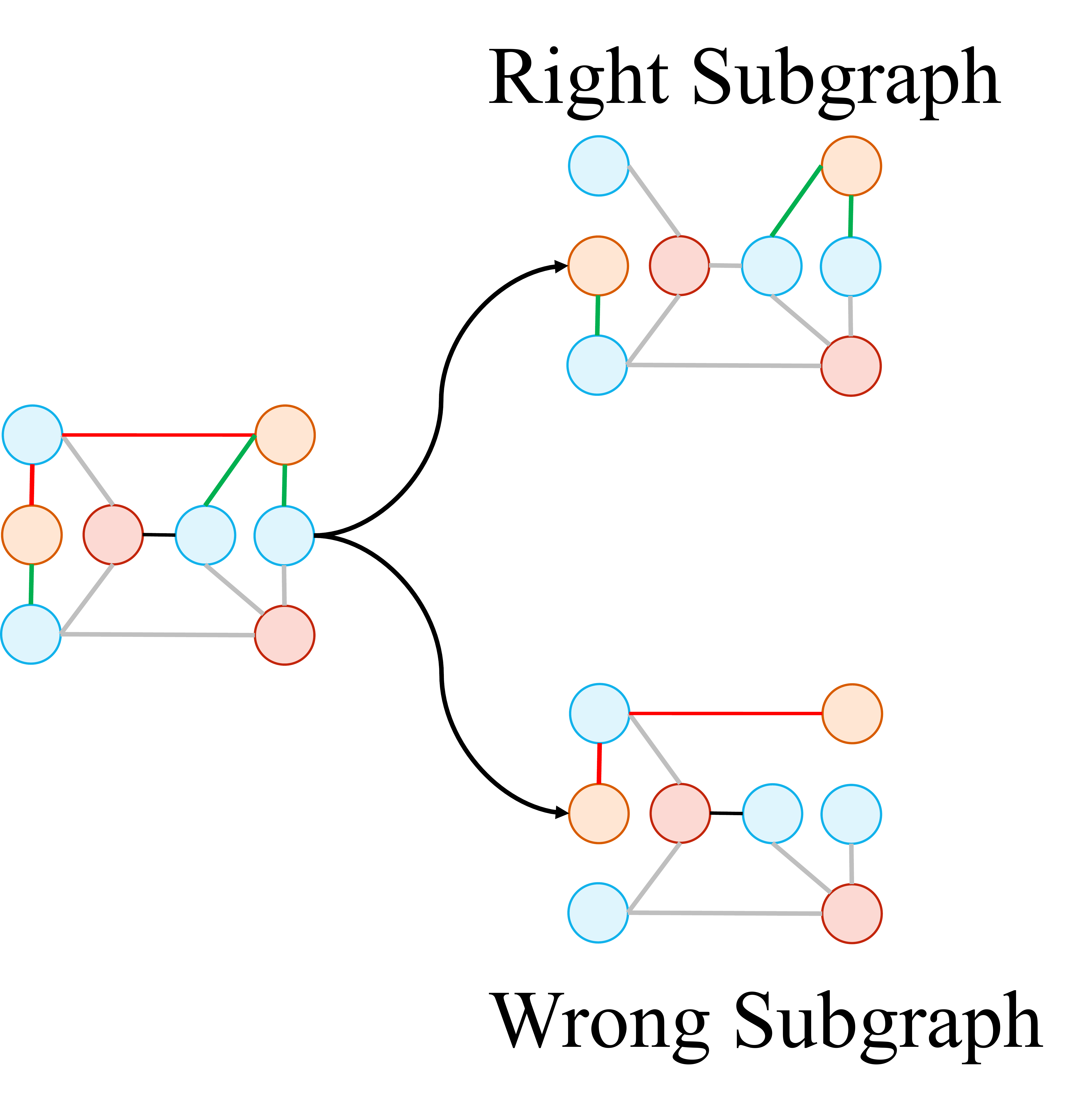}\\
    \end{minipage}
    \caption{(a) The proposed response graph. (b) Right-wrong decomposition.}
\label{fig:resg}
\end{figure}

\section{METHODOLOGY: The Proposed ORCDF}
This section introduces the proposed ORCDF. It starts by introducing the proposed novel response graph, then explores the response-aware graph convolution (RGC), a technique designed to capture the rich information embedded in the response graph. Following this, we introduce a consistency regularization loss function. We also discuss the model training and analyze model complexity. An overview of ORCDF is shown in Figure~\ref{fig:framework}.

\textbf{Response Graph.}
As illustrated in Figure~\ref{fig:resg}(a), focusing on responses, the response graph (ResG), denoted as $\mathcal{G}=(\mathcal{V}, \mathcal{E})$, comprises three types of nodes and edges. $\mathcal{V}= S \cup E \cup C$ involves students, exercises, and concepts, $\mathcal{E}$ involves interactions between $S$ and $E$ (i.e., ``Right''), $S$ and $E$ (i.e., ``Wrong''), $E$ and $C$ (i.e., ``Related''). Notably, we incorporate the response signal on exercises as the edge types between students' nodes and exercises' nodes. Next, we will introduce how to capture the fruitful response signal information.

\subsection{Response-aware Graph Convolution}
\textbf{Construct Embeddings.} In CD, the primary data elements are
response logs and the $\mathbf{Q}$. It is crucial to deconstruct these complex logs into their fundamental components: students, exercises, and concepts. We encode them with trainable embeddings $\mathbf{H}_s \in \mathbb{R}^{N \times d}, \mathbf{H}_e \in \mathbb{R}^{M \times d}, \mathbf{H}_c \in \mathbb{R}^{Z \times d}$. For instance, $\mathbf{h}_{s_i} \in \mathbf{R}^{1 \times d}$ denotes the row vector of the $i$-th student. To facilitate subsequent convolution processes, we stack the aforementioned embeddings to form $\mathbf{H}^{(0)} \in \mathbb{R}^{(N+M+Z) \times d}$.

\textbf{Right-Wrong Decomposition.} In the ResG, there are two types
of response signals existing between student nodes and exercise
nodes, as shown in Figure~\ref{fig:resg}(a). To better explore the impact of different response signals on learning Mas, we intuitively decompose
the response graph into a right subgraph and a wrong subgraph.
From the perspective of adjacency matrix, this involves splitting
the interaction matrix $\mathbf{I}$ into $\mathbf{I}_{\text{right}}$ (1 represents right, 0 represents others) and $\mathbf{I}_{\text{wrong}}$ (1 represents wrong, 0 represents others). \textit{For brevity, in the following sections, we will denote ``R'' for right and ``W'' for wrong.} Then we construct the right and wrong subgraphs (i.e, $\mathbf{A}_{\text{R}}, \mathbf{A}_{\text{W}}$ as expressed by Eq.~\eqref{eq:des}:
\begin{equation}
    \mathbf{A}_{\text{R}}=\left(\begin{array}{ccc}
\mathbf{O} & \mathbf{I}_{\text{R}} & \mathbf{O}  \\
\mathbf{I}_{\text{R}}^{\top} & \mathbf{O} & \mathbf{Q} \\
\mathbf{O} & \mathbf{Q}^{\top} & \mathbf{O}
\end{array}\right), \quad \mathbf{A}_{\text{W}}=\left(\begin{array}{ccc}
\mathbf{O} & \mathbf{I}_{\text{W}} & \mathbf{O} \\
\mathbf{I}_{\text{W}}^{\top} & \mathbf{O} & \mathbf{Q} \\
\mathbf{O} & \mathbf{Q}^{\top} & \mathbf{O}
\end{array}\right)  \label{eq:des}\,.
\end{equation}

In the ResG, the neighbors of a specific exercise node may include students who either answer the exercise right or wrong. However, after disentangling such response signals in the ResG, in each subgraph, the neighbors of a particular exercise node will only consist of students who displayed the same response signals. For example, if both \( s_1 \) and \( s_2 \) are connected to \( e_1 \), indicating that they both answer \( e_1 \) correctly, there may be some shared information explaining why they both got it right. Such crucial response signals will be propagated during the message-passing mechanism by GCN. This process enables a deeper understanding of the nuances in student responses. In the following part, we will introduce a novel graph convolution approach tailored to capture the information from the two disentangled subgraphs in CD.

\textbf{Embedding Propagation.} Considering that in CD, the features of students, exercises, and knowledge concepts are quite simple, consisting only of IDs, we draw inspiration from [7]. As a result, we eliminate linear transformations and nonlinear activation functions, opting to use only the fundamental components of GCN. Hence, the graph embedding propagation layer is designed with the following matrix form
\begin{equation}\label{eq:conv}
\mathbf{H}^{(l)} = \hat{\mathbf{A}}\mathbf{H}^{(l-1)}, \quad \hat{\mathbf{A}} = \left( \mathbf{D}^{-\frac{1}{2}} \hat{\mathbf{A}} \mathbf{D}^{-\frac{1}{2}} \right),
\end{equation}
where \( \mathbf{A} \) can be \( \mathbf{A}_\text{R} \) or \( \mathbf{A}_\text{W} \). The degree matrix \( \mathbf{D} \) is a diagonal matrix with size \( (N + M + Z) \times (N + M + Z) \), where each entry \( \mathbf{D}_{ii} \) representing the number of non-zero entries in the $i$-th row vector of the matrix \( \mathbf{A} \). Using Eq.~\eqref{eq:conv}, we can obtain the convolution outcomes from the \( l \)-th layer of the disentangled subgraphs, specifically \( \mathbf{H}^{(l)}_\text{R} \) and \( \mathbf{H}^{(l)}_\text{W} \). However, right and wrong represent completely opposite response signals, it may be inappropriate to directly plus the results obtained from convolutions performed on the two subgraphs. A sophisticated function capable of aggregating these two types of information is necessary, as the interaction mechanisms between students and exercises are quite intricate. It can be expressed as
\begin{equation}
\mathbf{H}^{(l)}_F = \phi(\mathbf{H}^{(l)}_\text{R} \mathbf{W}_{\text{rc}} + \mathbf{H}^{(l)}_\text{W} \mathbf{W}_{\text{wc}})\,,
\end{equation}
where \( \mathbf{H}^{(l)}_F \) denotes the final representation of the \( l \)-th RGC layer and \( \phi \) denotes arbitrary nonlinear activate function. $\mathbf{W}_{\text{rc}}, \mathbf{W}_{\text{wc}} \in \mathbb{R}^{d \times d}$ are trainable parameters. Intuitively, \( \mathbf{H}^{(l)}_\text{R} \mathbf{W}_{\text{rc}} \) denotes the right channel which obtain the semantic information from right signal. Conversely, \( \mathbf{H}^{(l)}_\text{W} \mathbf{W}_{\text{wc}} \) represents the opposite. The ultimate embedding \( \mathbf{H}_F \) is calculated using a mean pooling operation on the outcomes from each layer of the RGC which can be expressed as
\begin{equation}\label{eq:mean}
\mathbf{H} = \frac{1}{1 + L} (\mathbf{H}^{(0)}_F + \mathbf{H}^{(1)}_F + \ldots + \mathbf{H}^{(L)}_F)\,.
\end{equation}

\textbf{Discussion.} Here, we explain why RGC can alleviate the oversmoothing issue in existing CDMs. Notably, since our goal is to alleviate the oversmoothing issue in CDMs, and given that shallow layers of RGC already achieve satisfactory experimental results, the oversmoothing issue caused by deep GNN is not addressed in this work and can be considered for future research. First, we analyse the NCDM which directly set Mas as \( \mathbf{H}^{(0)} \in \mathbb{R}^{N \times d} \) and \( d = Z \). Firstly, the MND of any two students \( s_1 \) and \( s_2 \) in original NCDM is calculated as
\begin{equation}\label{mnd:ncdm}
\text{MND}_{s_1, s_2} = \|\mathbf{Mas}_{s_1} - \mathbf{Mas}_{s_2}\|_2^2 = \|\mathbf{H}^{(0)}_{s_1} - \mathbf{H}^{(0)}_{s_2}\|_2^2\,.
\end{equation}

Clearly, the difference between the learned Mas of \( s_1 \) and \( s_2 \) in NCDM reflects the disparity in their individual information. Here, we will use a one-layer RGC incorporated with NCDM as an example. For brevity, we will omit all normalization coefficients, biases and retain only the key components. The MND of students \( s_1 \) and \( s_2 \) after utilizing RGC is calculated as
\begin{equation}
\text{MND}_{s_1, s_2} = \|\mathbf{Mas}_{s_1} - \mathbf{Mas}_{s_2}\|_2^2 = \|\mathbf{H}_{s_1} - \mathbf{H}_{s_2}\|_2^2\,.
\end{equation}

Via Eq.~\eqref{eq:conv}, we can derive that \( \mathbf{H}^{(1)}_{s_1} (\text{R}) = \sum_{e_j \in \mathcal{N}^\text{R}(s_1)} \mathbf{H}_{e_j} \mathbf{W}_\text{rc} \) where \( e_j \in \mathcal{N}^\text{R}(s_1) \) represents the \( j \)-th exercise \( s_1 \) practiced correctly and is also the neighbor of \( s_1 \) in the right subgraph. Consequently, the term \( \mathbf{H}^{(1)}_{s_1}(\text{R}) \) represents the normalized summation exercises where \( s_1 \) practiced correctly. Similarly, we can derive \( \mathbf{H}^{(1)}_{s_2} (\text{R}), \mathbf{H}^{(1)}_{s_1} (\text{W}) \) and \( \mathbf{H}^{(1)}_{s_2}(\text{W}) \) following the same logic. Finally, Via Eq.~\eqref{eq:conv} and Eq.~\eqref{eq:mean}, the \( \|\mathbf{H}_{s_1} - \mathbf{H}_{s_2}\|_2^2 \) can be calculated as \( \frac{1}{2}(\|\mathbf{H}^{(0)}_{s_1} - \mathbf{H}^{(0)}_{s_2} + \mathbf{H}^{(1)}_{s_1}(\text{R}) - \mathbf{H}^{(1)}_{s_2}(\text{R}) + \mathbf{H}^{(1)}_{s_1}(\text{W}) - \mathbf{H}^{(1)}_{s_2} (\text{W})\|_2^2) \). Consequently, we can have the following observations:

$\bullet$  The first term \( \mathbf{H}^{(0)}_{s_1} - \mathbf{H}^{(0)}_{s_2} \) is the same as Eq.~\eqref{mnd:ncdm} which reflects the disparity of individual information of \( s_1 \) and \( s_2 \).

$\bullet$  The second term \( \mathbf{H}^{(1)}_{s_1} (\text{R}) - \mathbf{H}^{(1)}_{s_2} (\text{R}) + \mathbf{H}^{(1)}_{s_1} (\text{W}) - \mathbf{H}^{(1)}_{s_2} (\text{W}) \) captures the difference in the exercises that students \( s_1 \) and \( s_2 \) practiced correctly and incorrectly.

$\bullet$  The final \( \text{MND}_{s_1, s_2} \) of RA-NCDM is the mean of the first and second terms which can be interpreted as a comparison of the differences between students from the aforementioned perspectives.

For instance, if both \( s_1 \) and \( s_2 \) have similar accuracy in their exercises, the first term will be quite small due to the monotonicity assumption in CD. However, if the exercises attempted by \( s_1 \) are more challenging compared to those of \( s_2 \), the second term will capture this disparity and consequently increase the final difference between \( s_1 \) and \( s_2 \). This suggests that the RGC is capable of capturing the differences in the exercises practiced by students, resulting in more distinctive Mas for each student.

Notably, as the number of RGC layers increases, the perspectives for considering student differences also multiply. For instance, a two-layer RGC would further compare the differences with other students who have similar exercise performance as the current student. Therefore, by incorporating RGC, CDMs can assess student differences from multiple angles, thereby mitigating the oversmoothing issue. We will validate this conclusion in our ablation study in Section~\ref{exp:ablation} and give visualizations of the learned Mas by T-SNE~\cite{tsne} in Section~\ref{sec:stu}.

\subsection{Consistency Regularization Loss}
After the graph convolution by multiple RGC layers, we can get the final representation \( \mathbf{H} \) via Eq.~\eqref{eq:mean}. However, as we disentangle the response signal and capture student differences from various perspectives, it may exacerbate the notorious impact of the guess and slip problem on CDMs~\cite{fuzzycdm, Yang2023Mcd}. Previous methods, as referenced in~\cite{DINA}, model the guess and slip probabilities for each exercise as fixed parameters. Evidently, this approach is somewhat brute-force and might overlook the individual impact of students. This is because the probability of guessing or slipping is likely to vary for each person across different exercises. Contrary to the aforementioned methods, in this paper, we treat guess and slip as noise edges within the ResG. Specifically, we flip the student-exercise edge type (i.e., from R to W or W to R) with a probability \( p_f \) in the ResG. This noised version of the ResG, where some edges are flipped, is referred to as the flipped ResG, as illustrated in the left part of Figure~\ref{fig:framework}. We aim for the representations derived from the original ResG and the flipped ResG to be similar, in order to ensure that the CDMs remain effective even when subject to the disturbances caused by guess and slip problem. It can be formulated as
\begin{equation}
\mathcal{L}_{\text{reg}} = - \sum_{s_a \in S} \log \left( \exp \left( \mathbf{h}^\prime_{s_a}\mathbf{h}^\top_{s_a} / \tau \right) \right)\,,
\end{equation}
where \( \mathbf{h}^\prime_{s_a} \) is the representation derived from flipped ResG, and \( \mathbf{h}_{s_a}^\prime \mathbf{h}^T_{s_a} \) denotes the similarity score the representation derived from the ResG and flipped ResG. \( \tau \) is the hyperparameter which controls the degree of smoothness utilized in various methods~\cite{Zhang2022Bc, Zhang2023Invcf, Zhang2023Info}.

\subsection{Model Training}
Given input embeddings, existing CDMs predict the performance of students practicing exercises, which can be formulated as
\begin{equation}
\hat{y}_{ij} = \mathcal{M}_{CD} (\mathbf{H}_{s_i}, \mathbf{H}_{e_j}, \mathbf{H}_c)\,,
\end{equation}
where \( \mathcal{M}_{CD}(\cdot) \) denotes the CDMs, and \( \mathbf{H} \) represents the input embedding that contains the representation of the student, exercises and concepts.

\textbf{Transformation Layer.} To facilitate the integration of ORCDF
with the majority of existing CDMs, we need to transform dimensions to suit the specific type of CDM in use. If the embedding size of CDMs is a latent dimension (e.g., KaNCD), we directly utilize $\mathbf{H}$ as the input embedding for incorporated CDMs. Otherwise (e.g., NCDM), we introduce a transformation layer which can be
formulated as
\begin{equation}
    \mathbf{H}_\text{t}= \mathbf{H}\mathbf{W}_\text{t}+\mathbf{b}_\text{t}\,,
\end{equation}
where \( \mathbf{H}_\text{t} \) will be employed as input embedding for incorporated CDMs and \( \mathbf{W}_\text{t} \in \mathbb{R}^{d \times Z} \), \( \mathbf{b}_\text{t} \in \mathbb{R}^{(N+M+Z) \times 1} \) are trainable parameters. As a result, unlike the previous RCD which sets \( d = Z \), we can choose \( d \) as a latent dimension (e.g., 64). This significantly reduces the time complexity of graph convolution, a point that will be further analyzed in the subsequent subsection.

\textbf{Joint Training.} The primary loss employed in CD task is to calculate the BCE loss between the model's predictions and the true response scores in a mini-batch. The aforementioned consistency regularization loss is incorporated jointly optimized with the CD task. The overall loss can be expressed as
\begin{equation}
\mathcal{L}_{\text{BCE}} = - \sum_{(s,e,r_{se}) \in T} \left[ r_{se} \log \hat{y}_{se} + (1 - r_{se}) \log (1 - \hat{y}_{se}) \right]\,,
\end{equation}
\begin{equation}
\mathcal{L} = \mathcal{L}_{\text{BCE}} + \lambda_{\text{reg}} \mathcal{L}_{\text{reg}}\,.
\end{equation}
\( \lambda_{\text{reg}} \) is a hyperparameter that governs the relative importance of the consistency regularization loss.

\subsection{Model Complexity Analysis}
Theoretically, we reveal that the graph convolution in ORCDF takes \( O(4|\mathcal{E}|Ld) \) time complexity. \( L \) denotes the number of RGC layers, and \( d \) denotes the dimension of embeddings. By leveraging the lightweight backbone and the transformation, our method is significantly lower in time complexity compared with the recent GNN-based approach RCD~\cite{RCD}. Specifically, RCD has the complexity of \( O(2|\mathcal{E}|LZ^2) \), where \( Z \) represents the number of concepts (\( d \ll Z \)). It suggests that ORCDF is more suitable for current online education scenario on ground of the increasing granularity of knowledge concepts. Indeed, ORCDF showcases a notable speed advantage, being up to \textit{\textbf{18 times}} faster than RCD on the Assist17 dataset (i.e., \( Z = 102 \)). This dataset is collected from ASSISTment online tutoring systems and extensively utilized CDMs~\cite{hiercdf}. This improvement comes along with enhanced performance and lower GPU memory usage. For detailed information, please refer to Appendix~\ref{appd:A}.

\section{Experiments}
In this section, we first describe four real-world datasets and evaluation metrics. Then, through extensive experiments, we aim to verify the superiority of ORCDF, which not only assists existing CDMs in mitigating the oversmoothing issue but also enhances the models' prediction performance and interpretability performance. To ensure the reliability and reproducibility of our experiments, they are independently repeated ten times with different seeds and our code is available at \url{https://github.com/lswhim/ORCDF}.

\begin{table}[!b]
\centering
\caption{Statistics of real-world datasets for experiments.}
\label{tab:datasets-stats}
\resizebox{0.85\linewidth}{!}{\begin{tabular}{l|cccc}
\midrule
Datasets & Assist17 & EdNet-1 & Junyi & XES3G5M \\
\midrule
\#Students & 1709 & 1776 & 10000 & 4000 \\
\#Exercises & 3162 & 11925 & 734 & 7191 \\
\#Knowledge Concepts & 102 & 189 & 734 & 832 \\
\#Response Logs & 390,311 & 616,193 & 408,057 & 1,174,514 \\
Sparsity & 0.072 & 0.029 & 0.055 & 0.04 \\
Average Correct Rate & 0.815 & 0.662 & 0.687 & 0.799 \\
$\mathbf{Q}$ Density & 1.22 & 2.25 & 1.0 & 1.16 \\
\bottomrule
\end{tabular}}
\end{table}

\subsection{Experimental Settings}
\textbf{Datasets Description.} The experiments are conducted using four real-world datasets: Assist17, EdNet-1, Junyi, and XES3G5M. The Assist17 dataset is provided by the ASSISTment web-based online tutoring systems~\cite{Assist0910} and are widely used for cognitive diagnosis tasks~\cite{NCDM}. EdNet-1~\cite{ednet-1} is the dataset of all student-system interaction collected over 2 years by Santa, a multi-platform AI web-based tutoring service with more than 780K users in Korea. Junyi~\cite{junyi} is an online math practice log dataset offered by Junyi Academy. XES3G5M~\cite{liu2023XES3G5M} is a knowledge tracing benchmark dataset with auxiliary information. For more detailed statistics on these four datasets, please cf. Table 1. Notably, ``Sparsity'' refers to the sparsity of the dataset, which is calculated as \( |T|/(|S||E|) \). ``Average Correct Rate'' represents the average score of students on exercises, and ``\( \mathbf{Q} \) Density'' indicates the average number of concepts per exercise.

\textbf{Evaluation Metrics.} To assess the efficacy of ORCDF, we utilize both score prediction, interpretability and oversmoothing metrics.

$\bullet$ \textbf{Score Prediction Metrics:} Assessing the effectiveness of CDMs poses difficulties owing to the absence of the true Mas. A common approach to address this challenge is to learn the Mas within the train data and then evaluate the models based on their learned Mas to predict students' performance on exercises in the test data. In line with prior CDM studies, we partition the response logs of students into train, valid and test data with 7:1:2 following the previous researches~\cite{NCDM} and assess CDMs' performance on the test data using classification metrics such as Area Under the ROC Curve (AUC), Accuracy (ACC). Crucially, we build the ResG solely based on the train data.

$\bullet$  \textbf{Interpretability Metric:} Diagnostic results are highly interpretable hold significant importance in CD. In this regard, we employ the degree of agreement (DOA), which is consistent with the approach used in~\cite{Liu2023Qccdm, Shen2024Scd, Liu2024Icdm}. The underlying intuition here is that, if \( s_a \) has a greater accuracy in answering exercises related to \( c_k \) than student \( s_b \), then the probability of \( s_a \) getting \( c_k \) should be greater than that of \( s_b \). Namely, \( \mathbf{Mas}_{s_a,c_k} > \mathbf{Mas}_{s_b,c_k} \). Details about DOA can be found in Appendix~\ref{appd:B}. Consistent with~\cite{hiercdf}, we compute the average DOA for the top 10 concepts with the highest number of response logs in Assist17, EdNet-1, Junyi and XES3G5M.

$\bullet$ \textbf{Oversmoothing Metric:} \label{eq:oversmoothing}We employ the proposed MND to measure the Mas learned by CDMs. In CD, since the Mas of students learned by CDMs with concept mastery pattern lies within the range of 0 to 1, we utilize the \( l_2 \) norm of the difference between two students' mastery level vectors to describe the disparity between them. It can be formulated as follows:
\begin{equation}
    \mathrm{MND}=\frac{1}{|S|} \frac{1}{|S|-1} \sum_{s_{u} \in S} \sum_{s_{v} \in S} \frac{\left\|\mathbf{Mas}_{s_{u}}-\mathbf{Mas}_{s_{v}}\right\|_{2}^{2}}{|C|}\,,
\end{equation}
where $S, C$ represent the set of students and knowledge concepts, respectively, and $\mathbf{Mas}_{s_u}$ stands for the learned Mas of student $s_u$ by CDMs. A larger MND value indicates greater difference in the Mas that learned by CDMs, implying that the oversmoothing issue is more adequately addressed. 

\textbf{Implementation Details.} For parameter initialization, we employ the Xavier~\cite{xavier},  and for optimization purposes, Adam~\cite{adam} is adopted. For fair comparison, the embedding size is uniformly
set to 32 for MIRT, KaNCD, and KSCD, and to $Z$ for NCDM and CDMFKC. The batch size is set as 4096 for all datasets. To regulate the impact of the regularization term, we adjust the flip ratio $p_f$ within the range $\{0.05, 0.15, 0.1, 0.2\}$,  $\lambda_{\text{reg}}$ within the range $\{10^{-4}, 10^{-3}, \dots, 10^{-1}\}$, $\tau$ within the range $\{0.1,0.5,1.0,3.0,5.0 \}$. Analysis regarding the aforementioned hyperparameters can be found in Section~\ref{exp:hyper} and Appendix~\ref{appd:C}. 

\subsection{Student Performance Prediction}
To showcase the effectiveness of ORCDF, we integrate it with various
CDMs, as described in the subsequent part.

$\bullet$ IRT~\cite{IRTA} is a classic model of latent factor CDMs, which uses
one dimension $\theta$ to model Mas and utilize logistic function as IF to
predict the student score performance.

$\bullet$ MIRT~\cite{MIRT} is a representative model of latent factor CDM,
which uses multidimensional $\bm{\theta}$ to model Mas.

$\bullet$ NCDM~\cite{NCDM} is the first recent deep-learning based CDM which
utilizes MLP to replace the traditional manually designed IFs.

$\bullet$ CDMFKC~\cite{Li2022CDMFKC} employs a sophisticatedly designed neural network to model the impact of knowledge concepts on students' score performance.

$\bullet$ KSCD~\cite{kscd} also delves into the implicit relationships among
knowledge concepts and employs a concept-augmented IF.

$\bullet$ KaNCD~\cite{KANCD} is an enhanced version of NCDM, delving into
the implicit relationships among concepts to tackle the knowledge
coverage issue.

\begin{table*}[!htbp]
  \centering
  \caption{Overall student score prediction performance. ``OL'' stands for ``original'', referring to the original method, and ``OR'' denotes the proposed ORCDF enhancement applied to the original method. Within each method, the entry that exhibits the highest mean value is highlighted in bold. The standard deviation is not shown in the table since it is very small (less than 0.01). If the mean value significantly differs from the original method, passing a $t$-test with a significance level of 0.01, then we denote it with ``*'' at the corresponding position. ``-'' indicates that the model is not suitable of calculating this metric. ``OOM'' signifies out-of-memory occurring on a single NVIDIA 3090 GPU. All metrics are ideally larger for better results.}
    \resizebox{0.9\linewidth}{!}{\begin{tabular}{c|c|cc|cc|cc|cc|cc|cc}
    \toprule
    \multirow{2}[2]{*}{Dataset} & \multirow{2}[2]{*}{Metric (\%)} & \multicolumn{2}{c|}{IRT} & \multicolumn{2}{c|}{MIRT} & \multicolumn{2}{c|}{NCDM} & \multicolumn{2}{c|}{CDMFKC} & \multicolumn{2}{c|}{KSCD} & \multicolumn{2}{c}{KANCD} \\
          &       & OL & OR & OL    & OR    & OL    & OR    & OL    & OR    & OL    & OR    & OL    & OR \\
    \midrule
    \multirow{4}[2]{*}{Assist17} 
    & AUC   & 88.95 & $\textbf{89.60}^*$ & 91.42 & $\textbf{91.95}^*$ & 86.89 & $\textbf{89.94}^*$ & 87.30  & $\textbf{90.02}^*$ & 88.56 & $\textbf{89.68}^*$ & 88.56 & $\textbf{90.33}^*$ \\
          & ACC   & 86.11 & $\textbf{86.75}^*$ & 88.15 & $\textbf{88.51}^*$ & 84.56 & $\textbf{87.10}^*$ & 85.15 & $\textbf{87.2}^*$ & 86.14 & $\textbf{86.75}^*$ & 86.06 & $\textbf{87.56}^*$ \\
          & DOA   & -  & -  & -  & -  & 51.39 & $\textbf{66.76}^*$ & 54.69 & $\textbf{66.67}^*$ & 65.86 & $\textbf{68.05}^*$ & 62.86 & $\textbf{67.01}^*$ \\
          & MND   & -  & -  & -  & -  & 1.43  & $\textbf{7.57}^*$ & 4.64  & $\textbf{20.7}^*$ & 0.05  & $\textbf{2.21}^*$ & 3.51  & $\textbf{14.08}^*$ \\
    \midrule
    \multirow{4}[2]{*}{EdNet-1} & AUC   & 73.18 & $\textbf{74.56}^*$ & 74.41 & $\textbf{74.68}^*$ & 72.86 & $\textbf{74.81}^*$ & 73.05 & $\textbf{74.85}^*$ & 73.74 & $\textbf{74.66}^*$ & 74.42 & $\textbf{75.11}^*$ \\
          & ACC   & 70.89 & $\textbf{71.85}^*$ & 71.70  & $\textbf{71.89}^*$ & 70.68 & $\textbf{71.98}^*$ & 70.79 & $\textbf{71.95}^*$ & 71.42 & $\textbf{71.85}^*$ & 71.75 & $\textbf{72.07}^*$ \\
          & DOA   & -  & -  & -  & -  & 59.31 & $\textbf{64.29}^*$ & 60.45 & $\textbf{64.01}^*$ & 64.55 & $\textbf{65.07}^*$ & 63.02 & $\textbf{65.47}^*$ \\
          & MND   & -  & -  & -  & -  & 1.42  & $\textbf{4.29}^*$ & 0.82  & $\textbf{4.05}^*$ & 0.05  & $\textbf{2.45}^*$ & 5.48  & $\textbf{7.12}^*$ \\
    \midrule
    \multirow{4}[2]{*}{Junyi} & AUC   & 80.35 & $\textbf{81.46}^*$ & 80.87 & $\textbf{81.46}^*$ & 77.72 & $\textbf{81.44}^*$ & 78.27 & $\textbf{81.30}^*$ & \multicolumn{2}{c|}{\multirow{4}[2]{*}{OOM}} & 79.12 & $\textbf{81.72}^*$ \\
          & ACC   & 76.65 & $\textbf{77.52}^*$ & 77.28 & $\textbf{77.54}^*$ & 74.49 & $\textbf{77.59}^*$ & 74.95 & $\textbf{77.28}^*$ & \multicolumn{2}{c|}{} & 75.57 & $\textbf{77.71}^*$ \\
          & DOA   & -  & -  & -  & -  & 49.92 & $\textbf{58.19}^*$ & 49.92 & $\textbf{60.74}^*$ & \multicolumn{2}{c|}{} & 53.59 & $\textbf{60.85}^*$ \\
          & MND   & -  & -  & -  & -  & 0.51  & $\textbf{11.22}^*$ & 0.34  & $\textbf{17.18}^*$ & \multicolumn{2}{c|}{} & 2.86  & $\textbf{12.82}^*$ \\
    \midrule
    \multirow{4}[2]{*}{XES3G5M} & AUC   & 79.18 & $\textbf{80.13}^*$ & 80.43 & $\textbf{80.66}^*$ & 75.46 & $\textbf{80.22}^*$ & 74.15 & $\textbf{79.98}^*$ & \multicolumn{2}{c|}{\multirow{4}[2]{*}{OOM}} & 79.68 & $\textbf{80.41}^*$ \\
          & ACC   & 81.52 & $\textbf{82.51}^*$ & 82.31 & $\textbf{82.52}^*$ & 81.21 & $\textbf{82.49}^*$ & 80.17 & $\textbf{82.28}^*$ & \multicolumn{2}{c|}{} & 82.23 & $\textbf{82.44}^*$ \\
          & DOA   & -  & -  & -  & -  & 68.01 & $\textbf{73.93}^*$ & 69.03 & $\textbf{73.89}^*$ & \multicolumn{2}{c|}{} & 73.50  & $\textbf{73.62}^*$ \\
          & MND   & -  & -  & -  & -  & 1.04  & $\textbf{19.37}^*$ & 2.83  & $\textbf{35.26}^*$ & \multicolumn{2}{c|}{} & 6.43  & $\textbf{16.67}^*$ \\
    \bottomrule
    \end{tabular}
    
}
  \label{tab:pred}
\end{table*}

\textbf{Details.} To ensure fairness in comparison, we adhere to the
hyperparameter settings and IFs as specified in their original publications. IRT and MIRT are non-interpretable models, namely latent factor CDMs, the Mas they learn cannot be correlated directly with specific knowledge concepts. Therefore, they are not suitable for calculating DOA and MND. In Table 2, we use ``-'' to indicate this inapplicability. If CDMs signify out-of-memory on an NVIDIA 3090 GPU, we use the term ``OOM'' to denote this occurrence.

\textbf{Experimental Results.} The main observations are as follows. As shown in Table~\ref{tab:pred}, the proposed ORCDF consistently and substantially improves the MND values of all the base CDMs
across all datasets. This confirms the efficacy of ORCDF in alleviating
the oversmoothing problem. We will validate the improvement of downstream tasks resulting from mitigating the oversmoothing
issue in Section~\ref{exp:cat}. Besides, from Table~\ref{tab:pred}, ORCDF significantly benefits the base CDMs. It substantially enhances both the prediction performances of all the base CDMs across various datasets. It is validated that the
ORCDF effectively alleviates the oversmoothing issue without compromising the prediction performance of the CDMs. Notably, DOAs of CDMs have improved substantially. This suggests that learning Mas from multiple perspectives aligns more closely with the monotonicity assumption prevalent in educational measurement,
indicating enhanced interpretability performance.

\begin{figure}[!h]
\centering
\begin{minipage}{0.32\linewidth}\centering
    \includegraphics[width=\textwidth]{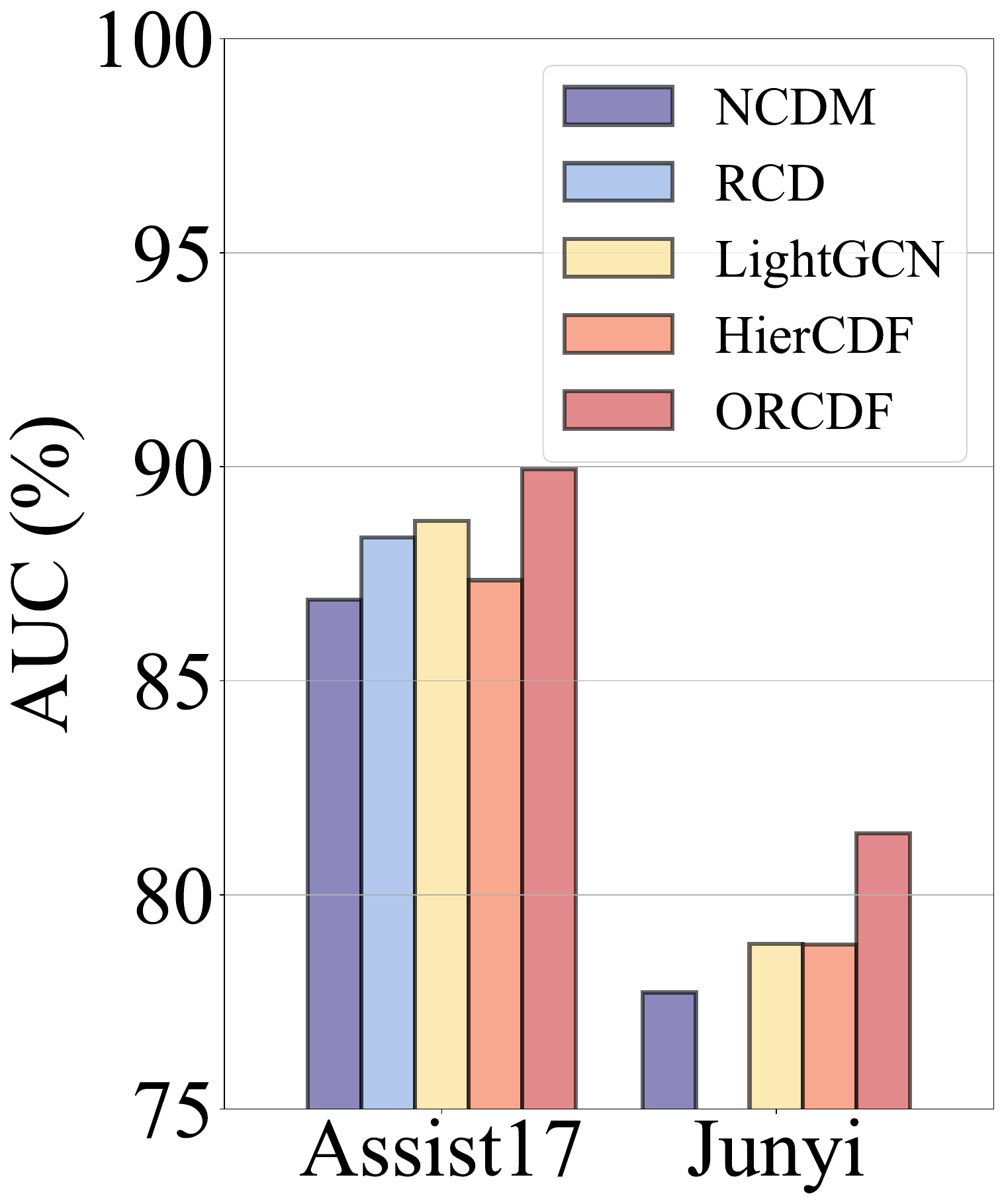}\\
\end{minipage}
\begin{minipage}{0.32\linewidth}\centering
    \includegraphics[width=\textwidth]{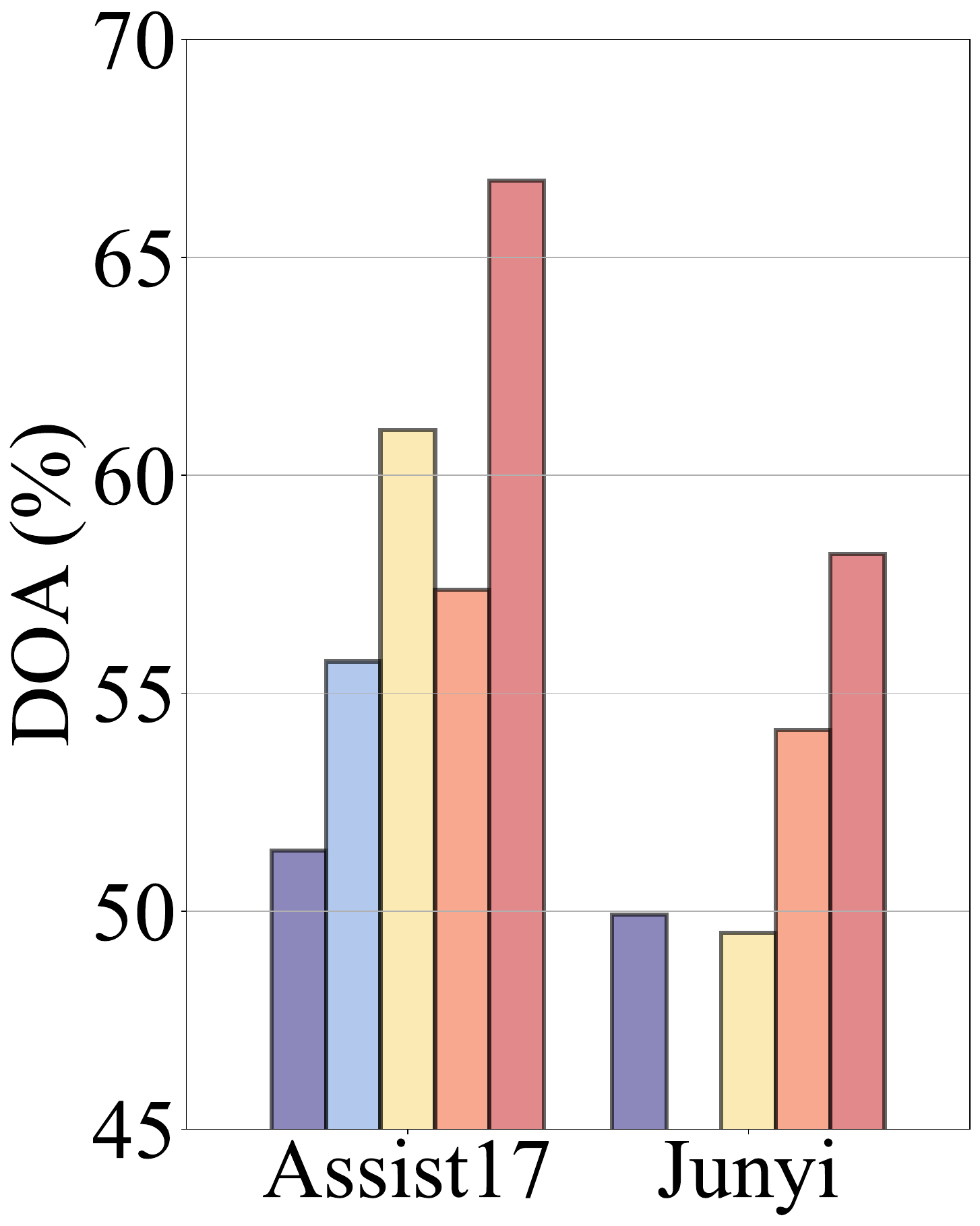}\\
\end{minipage}
\begin{minipage}{0.32\linewidth}\centering
    \includegraphics[width=\textwidth]{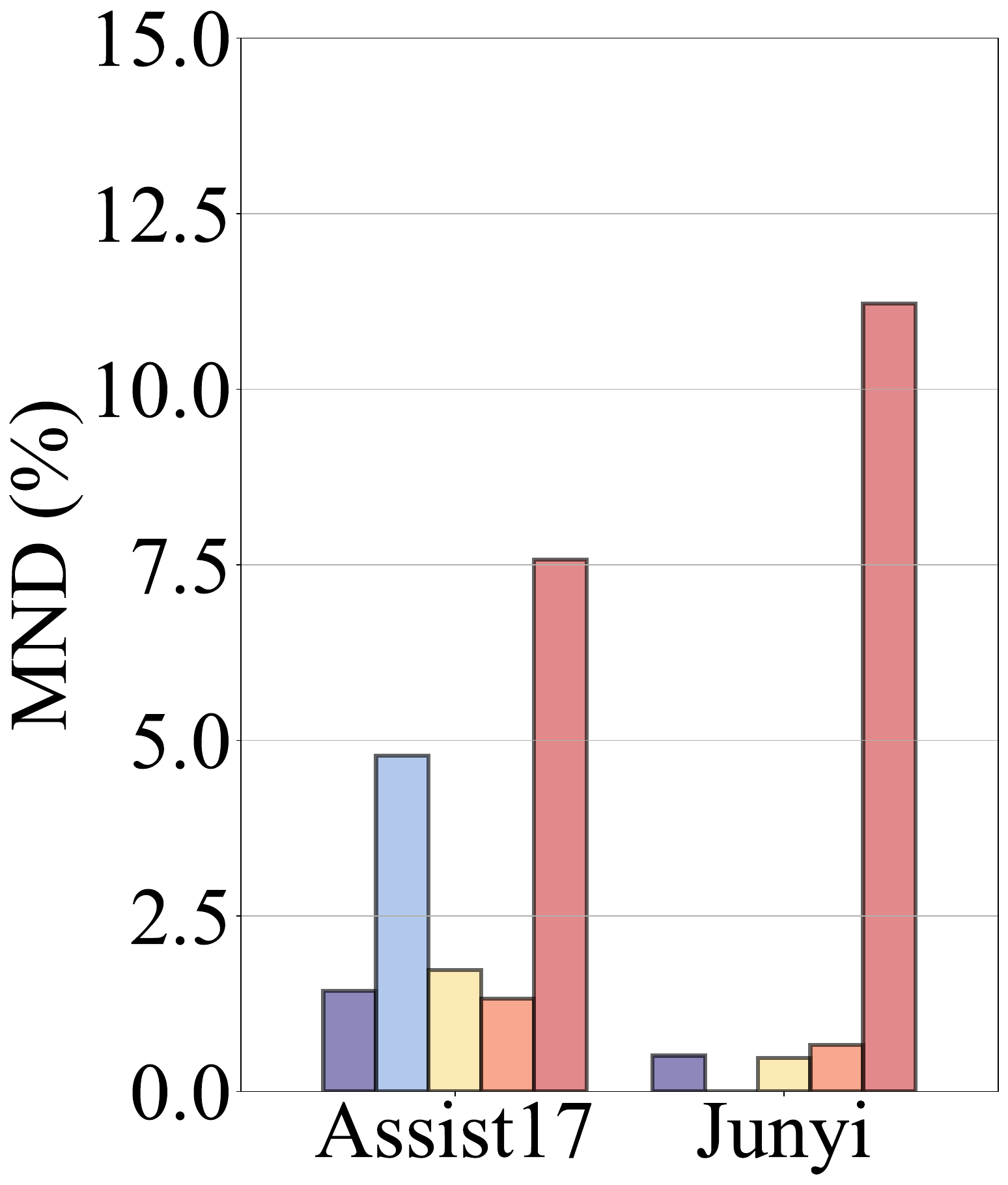}
\end{minipage}
\caption{Comparison with other related frameworks incorporating NCDM.}
\label{fig:compare}
\end{figure}

Furthermore, we conduct comparisons with other competitive frameworks including other graph-based methods or related approaches to further validate the effectiveness of the ORCDF. The results are in Figure~\ref{fig:compare}.

$\bullet$ RCD~\cite{RCD} is the first method to employ GAT in addressing
tasks within the field of CD. It uses standard GAT to delve into the
intricate relationships among students, exercises, and concepts.

$\bullet$ LightGCN~\cite{lightgcn} is a recent classic model that employs GCN
in collaborative filtering. Since LightGCN is a lightweight graph
neural network suitable for heterogeneous graphs with solely ID
as features, we chose it as the representative baseline.

$\bullet$ HierCDF~\cite{hiercdf} utilizes the Bayesian network to model the mastery pattern with directed acyclic graph of knowledge concepts.
For a fair comparison, we integrate the aforementioned methods with the NCDM and conduct the experiments on Assist17 and Junyi
under the same settings as previously described. Since RCD has already shown superiority over some heterogeneous graph representation learning methods such as HetG~\cite{zhang2019HGN}, and HAN~\cite{Wang2019HAN}, we do not include these methods in our comparative analysis. Since the RCD experiences OOM issue on Junyi, we do not report the results for this dataset. As shown in Figure~\ref{fig:compare}, ORCDF outperforms other chosen frameworks, whether they are specifically tailored for CD or other fields, in the task of predicting student performance. The superiority of ORCDF in terms of MND further confirms that learning Mas from multiple perspectives is beneficial for alleviating the oversmoothing issue.

\begin{table}[!htbp]
  \centering
  \caption{Ablation study of ORCDF. Details are as same as Table~\ref{tab:pred}.}
     \resizebox{0.82\linewidth}{!}{\begin{tabular}{c|c|cccc}
    \toprule
    \multirow{2}[2]{*}{Dataset} & \multirow{2}[2]{*}{Metric} & \multicolumn{4}{c}{NCDM} \\
          &       & OL    & OR-w/o-rgc & OR-w/o-reg & OR \\
    \midrule
    \multirow{4}[2]{*}{Assist17} & AUC   & 86.89 & 88.73 & 89.91 & \textbf{89.94} \\
          & ACC   & 84.56 & 86.19 & 87.07 & \textbf{87.10} \\
          & DOA   & 51.39 & 63.74 & 65.26 & \textbf{66.76} \\
          & MND   & 1.43  & 2.53  & 6.90   & \textbf{7.57} \\
    \midrule
    \multirow{4}[2]{*}{EdNet-1} & AUC   & 72.86 & 74.77 & 74.76 & \textbf{74.81} \\
          & ACC   & 70.86 & 71.94 & 71.86 & \textbf{71.98} \\
          & DOA   & 59.31 & 63.73 & 64.23 & \textbf{64.29} \\
          & MND   & 1.42  & 2.26  & 3.35  & \textbf{4.29} \\
    \multirow{4}[2]{*}{Junyi} & AUC   & 77.72 & 80.23 & 81.14 & \textbf{81.44} \\
          & ACC   & 74.49 & 76.52 & 77.22 & \textbf{77.59} \\
          & DOA   & 49.92 & 57.96 & 58.14 & \textbf{58.19} \\
          & MND   & 0.51  & 4.96  & 7.96  & \textbf{11.22} \\
    \midrule
    \multirow{4}[2]{*}{XES3G5M} & AUC   & 75.46 & 80.24 & 80.22 & \textbf{80.32} \\
          & ACC   & 81.21 & 82.46 & 82.46 & \textbf{82.49} \\
          & DOA   & 68.01 & 73.45 & 73.93 & \textbf{73.94} \\
          & MND   & 1.04  & 5.79  & 10.71 & \textbf{19.37} \\
    \bottomrule
    \end{tabular}}
  \label{tab:ab}
\end{table}

\subsection{Ablation Study}\label{exp:ablation}
In this subsection, we scrutinize and evaluate each key individual component of ORCDF to comprehend their respective impacts and significance on the overall performance of the model. The ablation analysis is conducted using the following three versions.

$\bullet$ OR-w/o-rgc: This ablation of ORCDF does not integrate the response-aware graph convolution. Instead, it directly perform convolution on the entire response graph without decomposition.

$\bullet$ OR-w/o-reg: This ablation of ORCDF does not utilize the proposed consistency regularization loss $\mathcal{L}_{\text{reg}}$.

$\bullet$ OL: It represents the base CDMs, which can be considered as the one without the inclusion of response-aware graph convolution and consistency regularization loss.

Due to space constraints, we only present the ablation study using OR-NCDM as an example. This choice is motivated by the fact that NCDM is often employed as a classic CDM in downstream tasks. It is worth noting that the results from incorporating other CDMs are generally similar.

\textbf{Experimental Results.} As indicated in Table~\ref{tab:ab}, the proposed method outperforms the other two versions, suggesting that each component plays a significant role in enhancing the model's overall effectiveness. OR-w/o-rgc performs significantly worse, further validating the superiority of the proposed RGC in capturing the information within the response graph. We empirically find that although the consistency regularization loss is designed to alleviate the guess and slip problem, it not only improves the prediction and interpretability performance but also
achieves a higher MND than the original version. This indicates that the guess and slip problem indeed exists in real-world scenarios, and addressing this problem is crucial for the effectiveness of CDMs.

\subsection{In-Depth Analysis of ORCDF's Advantages}
In this subsection, we analyze the proposed ORCDF from two perspectives: generalization performance and robustness performance. 

\textbf{Generalization Performance.} To assess the efficacy of ORCDF in addressing the generalization issue, we conduct experiments on
three datasets with varying test ratios $p_t = \{10\%, 20\%, 30\%, 40\%, 50\%\}$.
As $p_t$ increases which is consistent with~\cite{RCD}, the generalization ability of CDMs is tested more stringently. As depicted in Figure~\ref{fig:gener} of Appendix~\ref{appd:B}, with an increasing test ratio $p_t$, the number of response logs used for training decreases. However, OR-NCDM consistently outperforms NCDM, illustrating that ORCDF can provide more accurate diagnosis results with fewer student response records. This is particularly suitable for current online learning scenarios, where students often have limited response logs.
\begin{figure}[!h]
\centering
\begin{minipage}{0.24\linewidth}\centering
    \includegraphics[width=\textwidth]{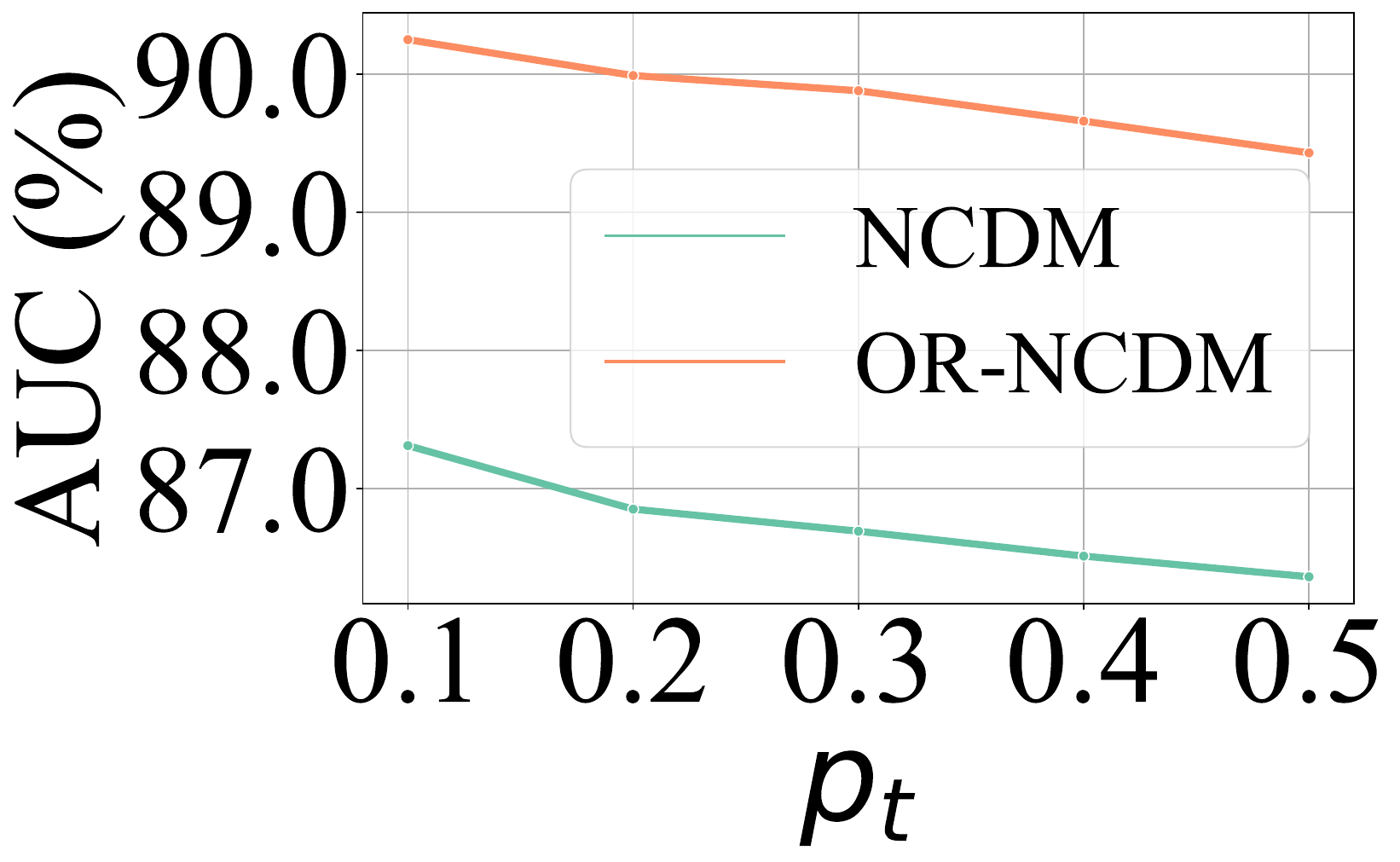}\\
\end{minipage}
\begin{minipage}{0.24\linewidth}\centering
    \includegraphics[width=\textwidth]{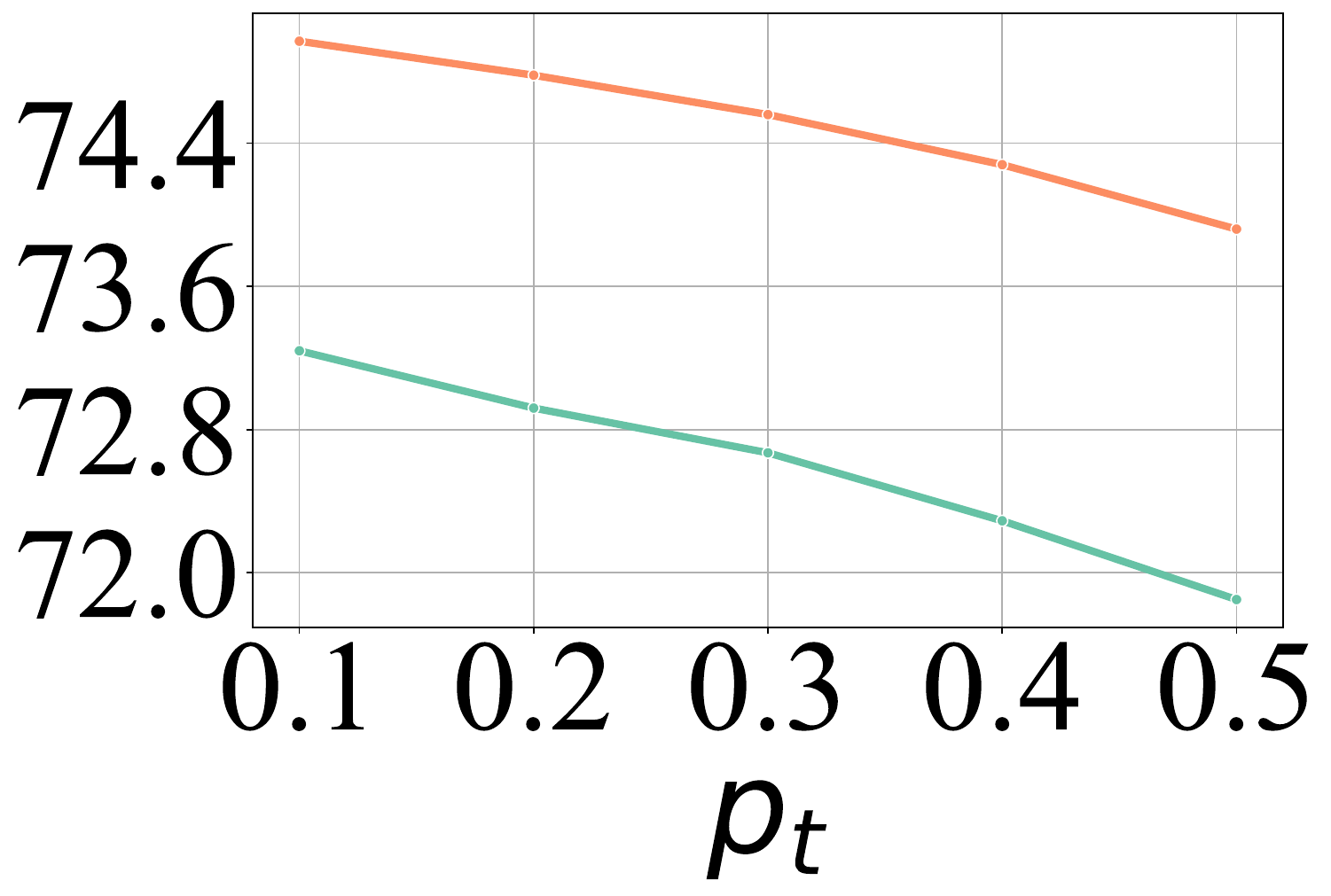}\\
\end{minipage}
\begin{minipage}{0.24\linewidth}\centering
    \includegraphics[width=\textwidth]{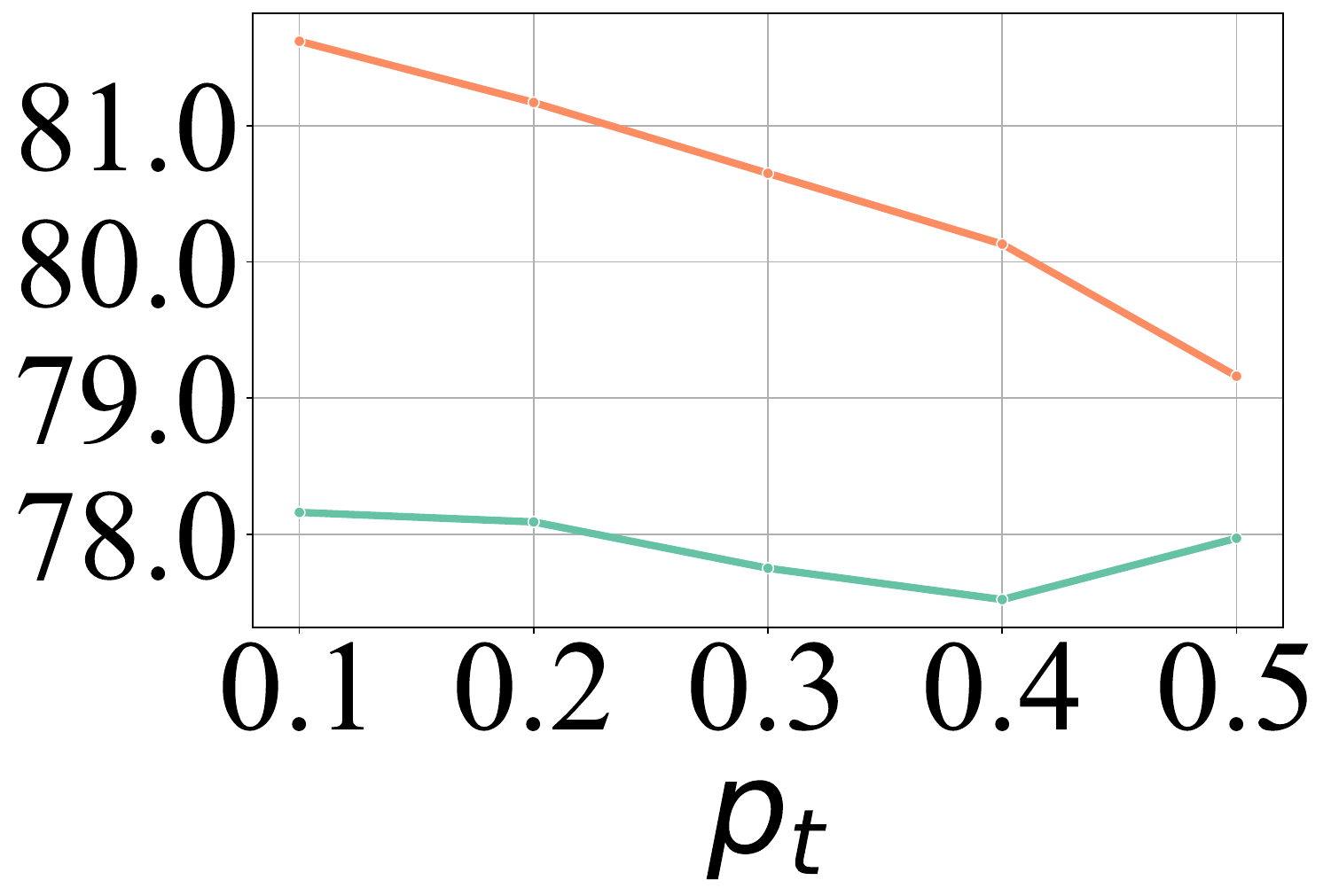}\\
\end{minipage}
\begin{minipage}{0.24\linewidth}\centering
    \includegraphics[width=\textwidth]{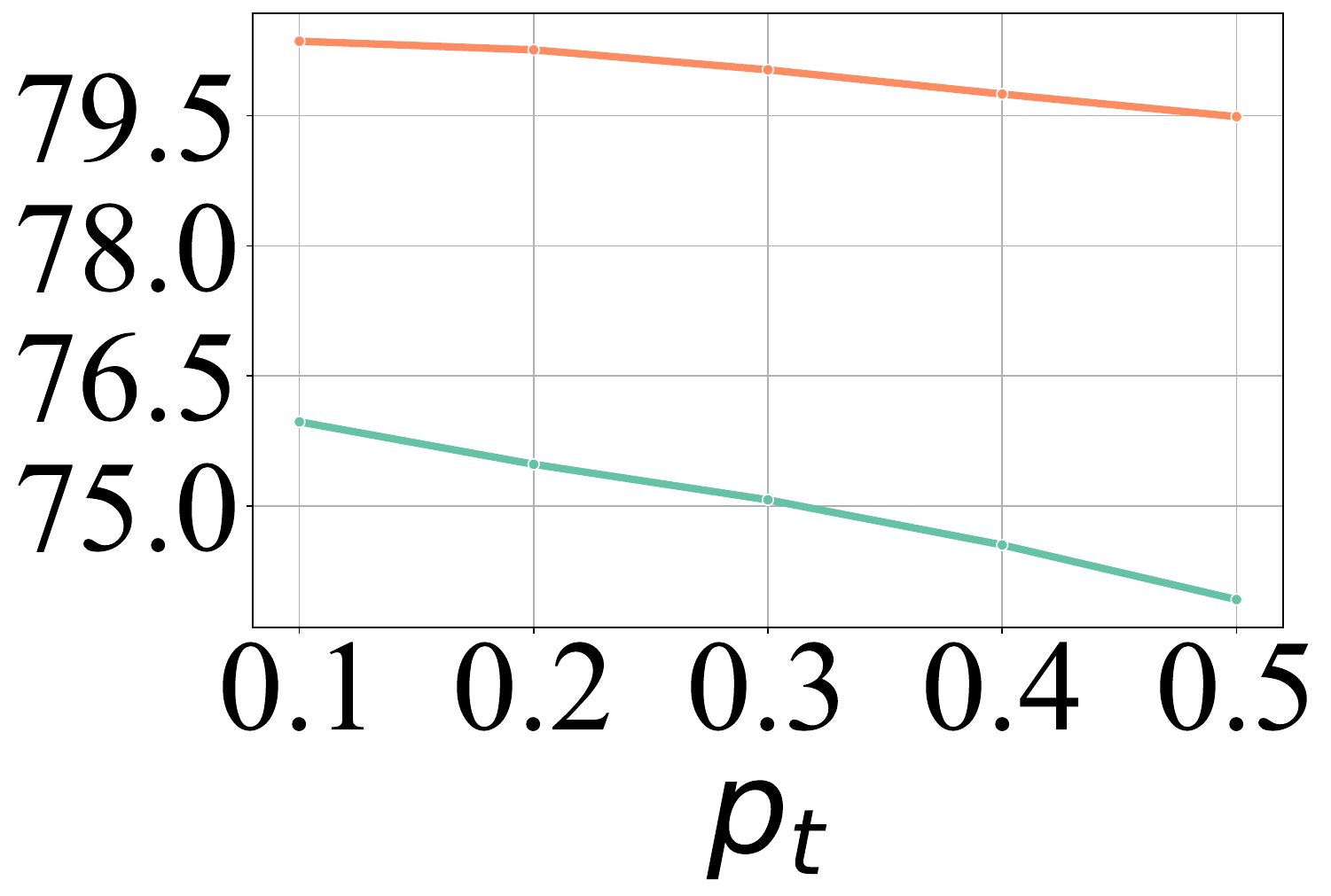}
\end{minipage}
\caption{Performance under different $p_t$ on four datasets.}
\label{fig:gener}
\end{figure}

\begin{figure}[!h]
\centering
\begin{minipage}{0.24\linewidth}\centering
    \includegraphics[width=\textwidth]{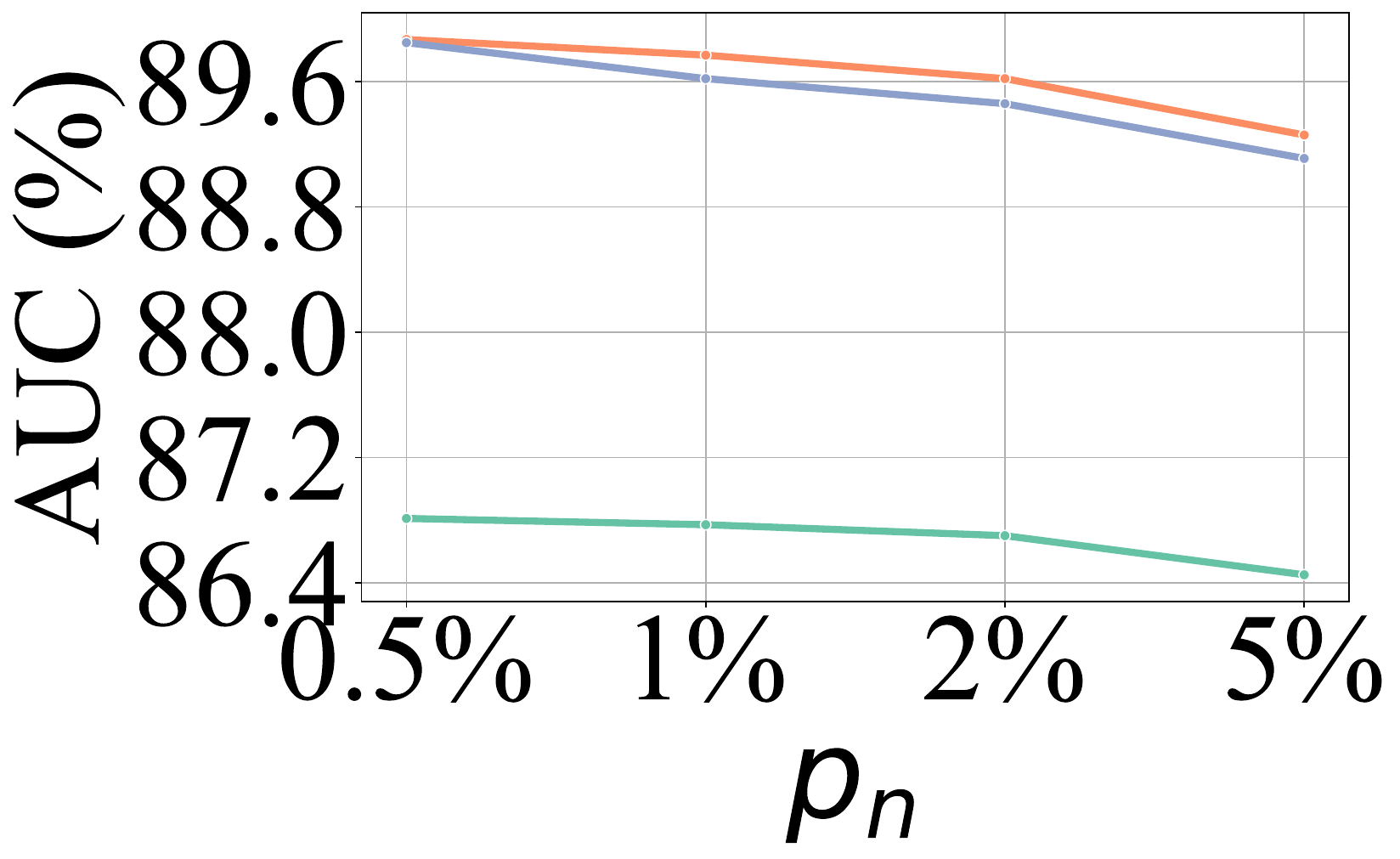}\\
\end{minipage}
\begin{minipage}{0.24\linewidth}\centering
    \includegraphics[width=\textwidth]{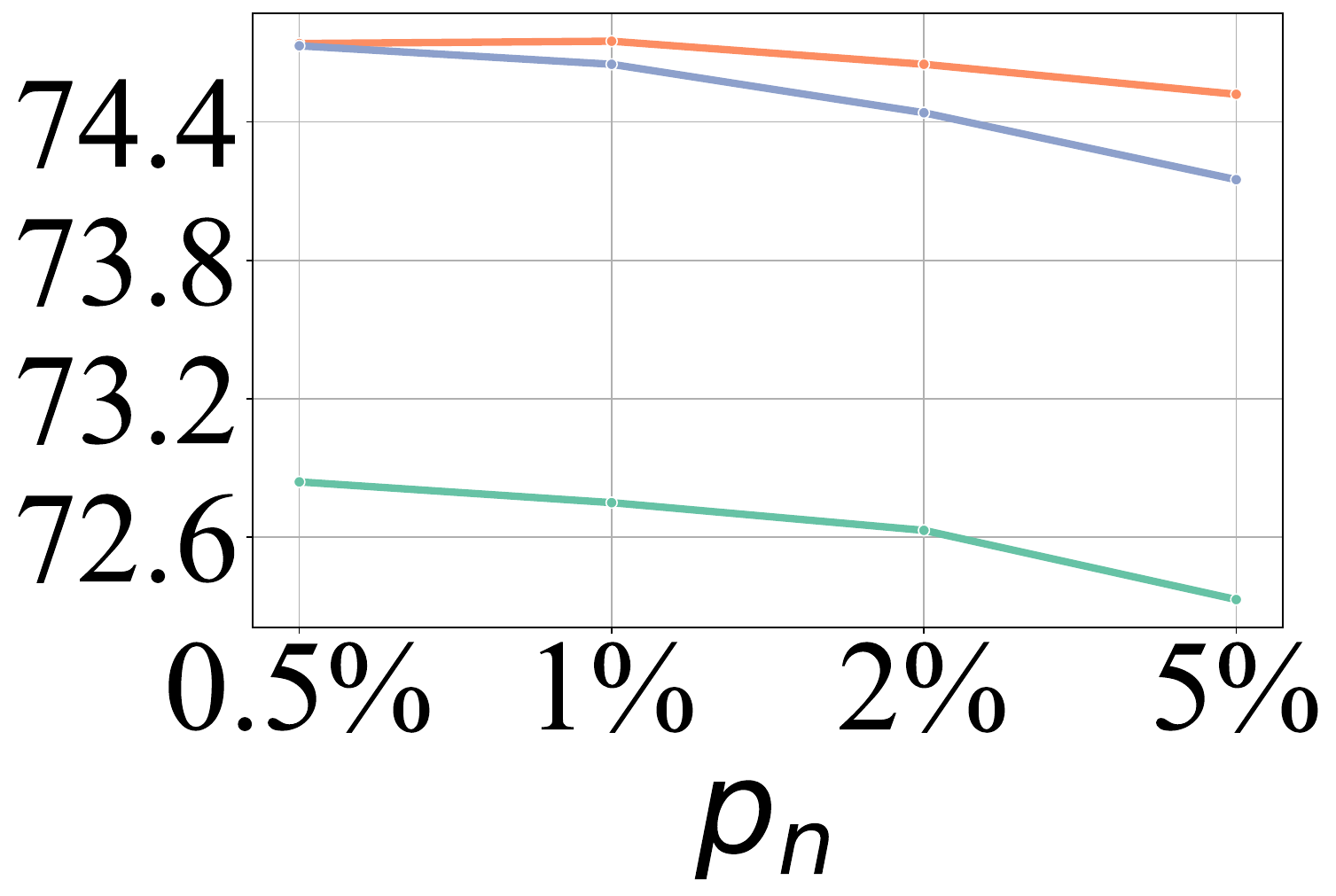}\\
\end{minipage}
\begin{minipage}{0.24\linewidth}\centering
    \includegraphics[width=\textwidth]{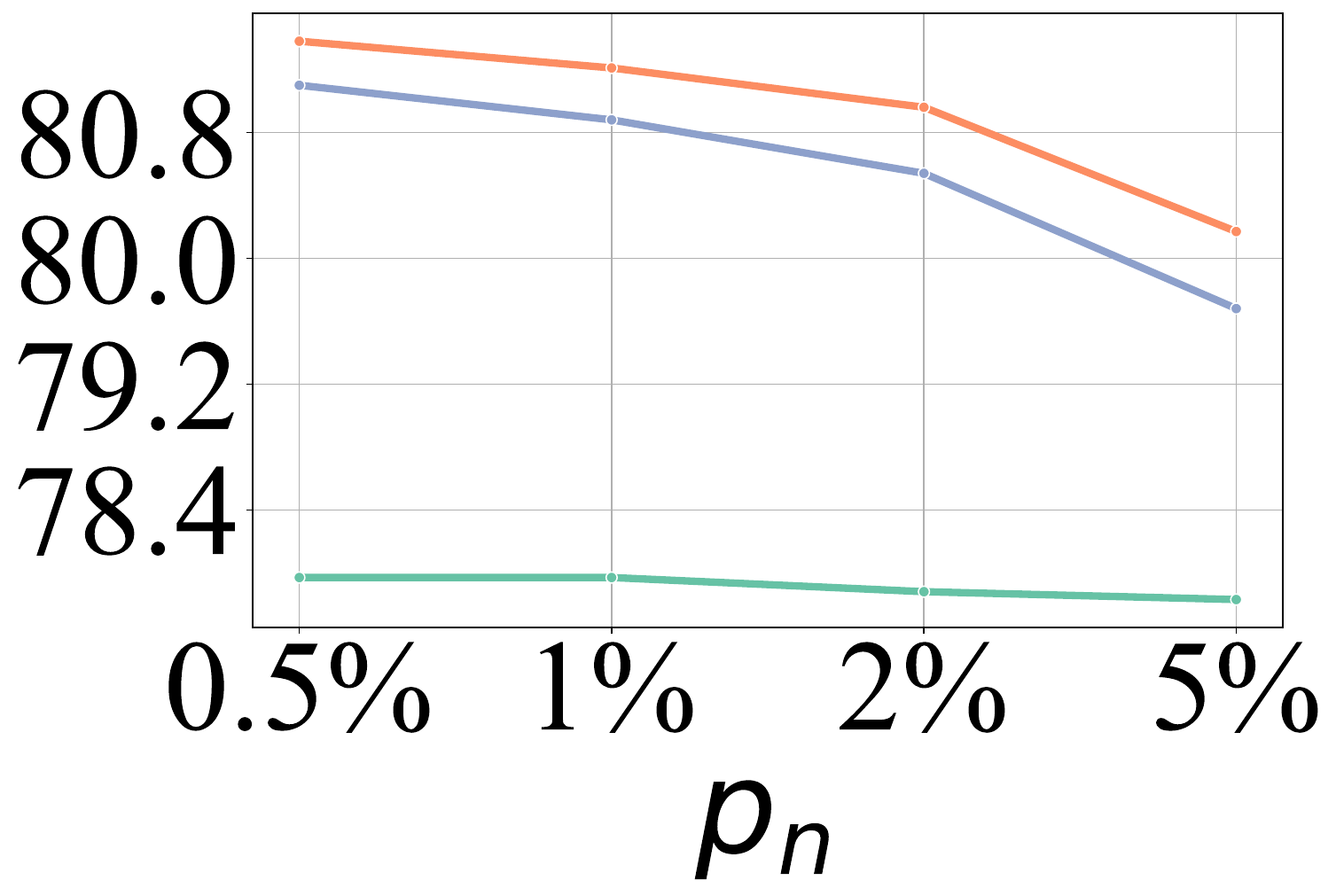}\\
\end{minipage}
\begin{minipage}{0.24\linewidth}\centering
    \includegraphics[width=\textwidth]{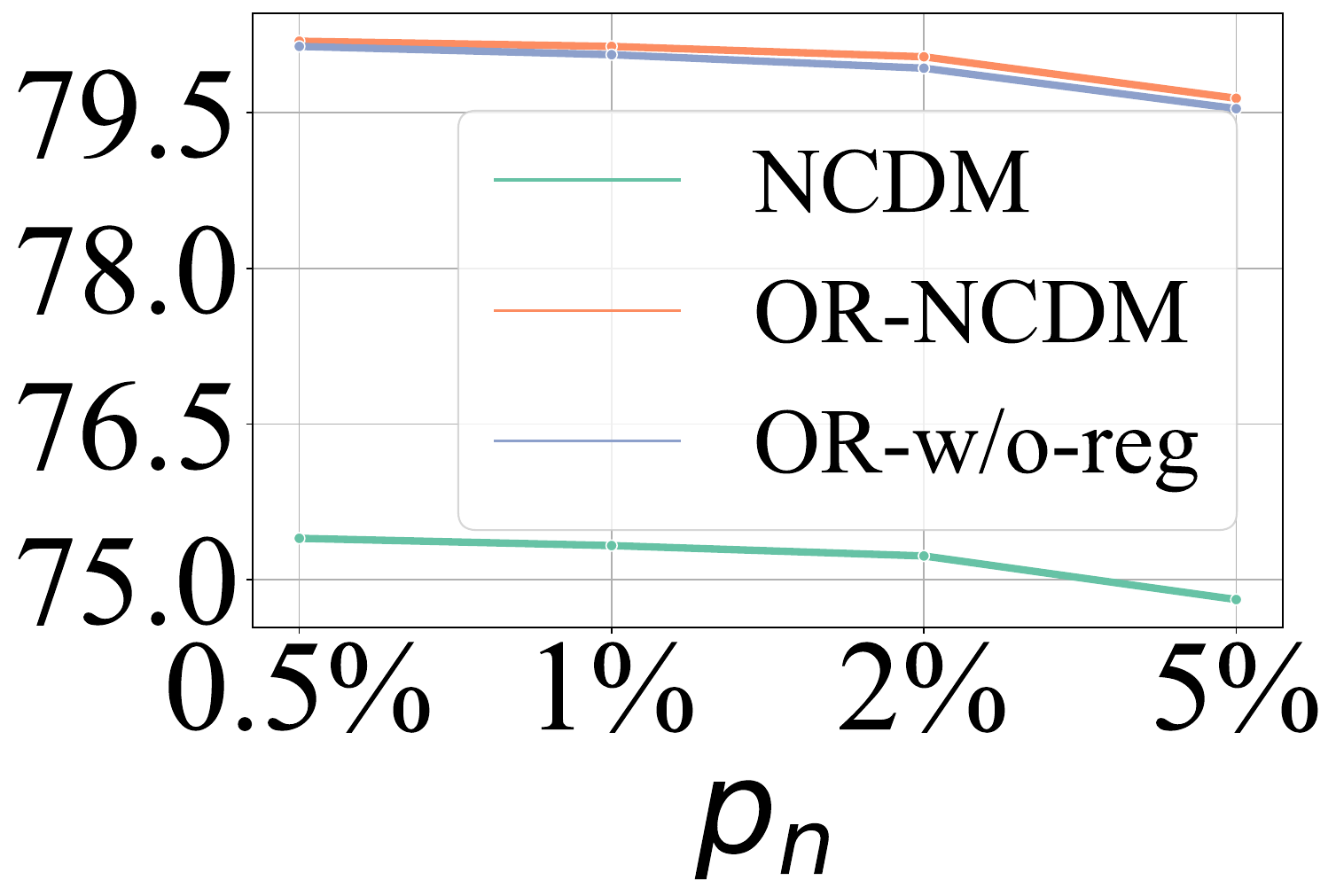}
\end{minipage}
\caption{Performance under different $p_n$ on four datasets.}
\label{fig:noise}
\end{figure}

\textbf{Robustness Performance.} Guess and slip problem~\cite{fuzzycdm, Yang2023Tkt} in CD can
significantly affect the accurate determination of students' Mas.
These noise interactions typically stem from two main factors:
guess and slip. To showcase the capacity of our proposed method,
ORCDF, in mitigating the issue of guess and slip problem, we can conceptually introduce noise into the training datasets while keeping
the test dataset unchanged. Specifically, to inject noise into the
train datasets, we can randomly select student responses and flip
them to the opposite. For example, correct responses can be flipped
to incorrect ($1 \rightarrow 0$) and vice versa ($0 \rightarrow 1$) at a certain noise ratio,
represented as $p_n$. As illustrated in Figure~\ref{fig:noise} of Appendix~\ref{appd:B}, as the noise ratio $p_n$ increases, the fact that OR-NCDM outperforms NCDM indicates its effectiveness in giving reasonable diagnosis result, especially when there is noises in the students' response logs. Notably, OR-NCDM shows a lesser performance drop than OR-w/o-rgc as the noise ratio increases, validating the effectiveness of our proposed loss function.

\begin{figure}[!h]
\centering
\begin{minipage}{0.44\linewidth}\centering
    \includegraphics[width=\textwidth]{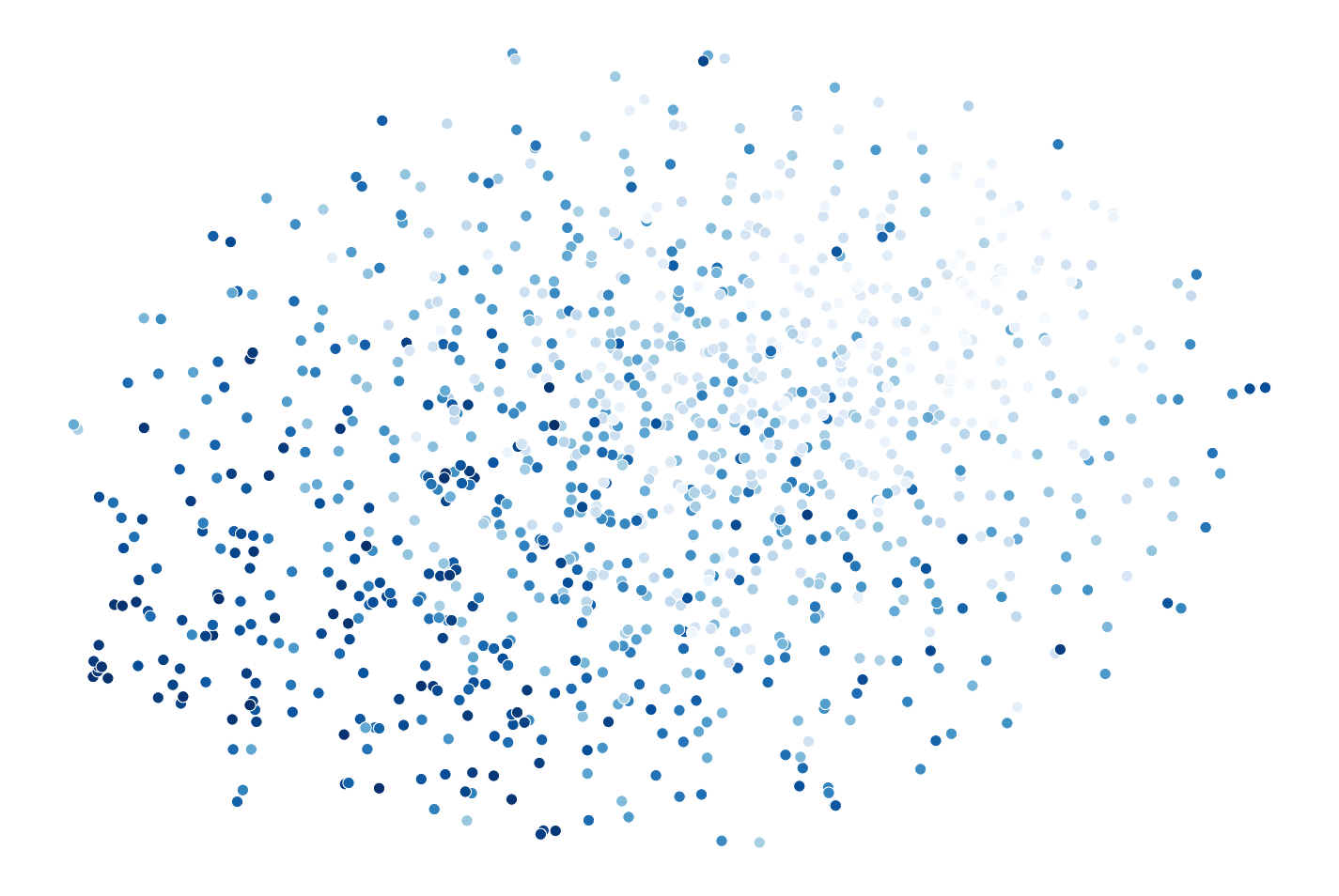}\\
    (a) NCDM
\end{minipage}
\begin{minipage}{0.44\linewidth}\centering
    \includegraphics[width=\textwidth]{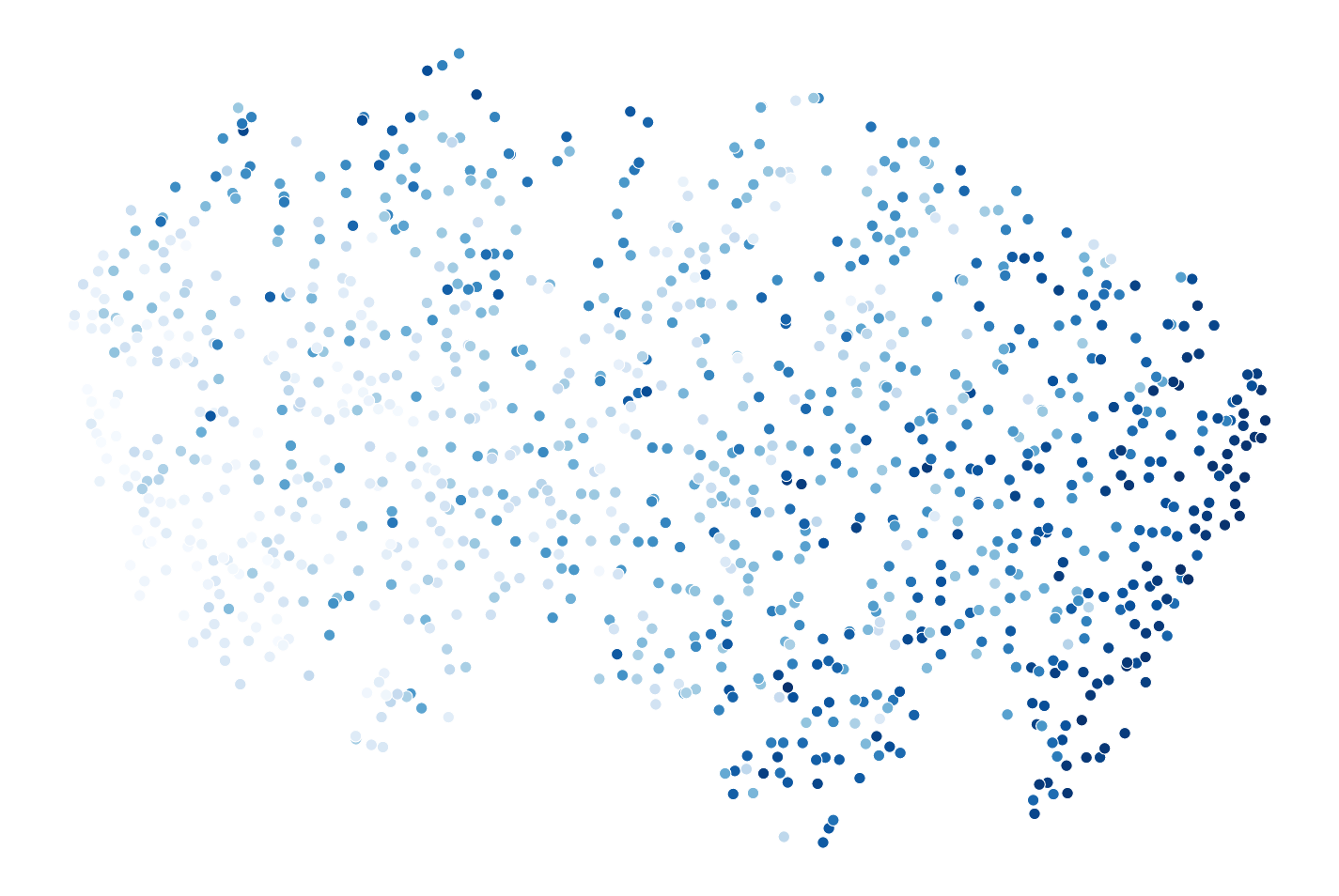}
    (b) OR-NCDM
\end{minipage}
\\
\begin{minipage}{0.44\linewidth}\centering
    \includegraphics[width=\textwidth]{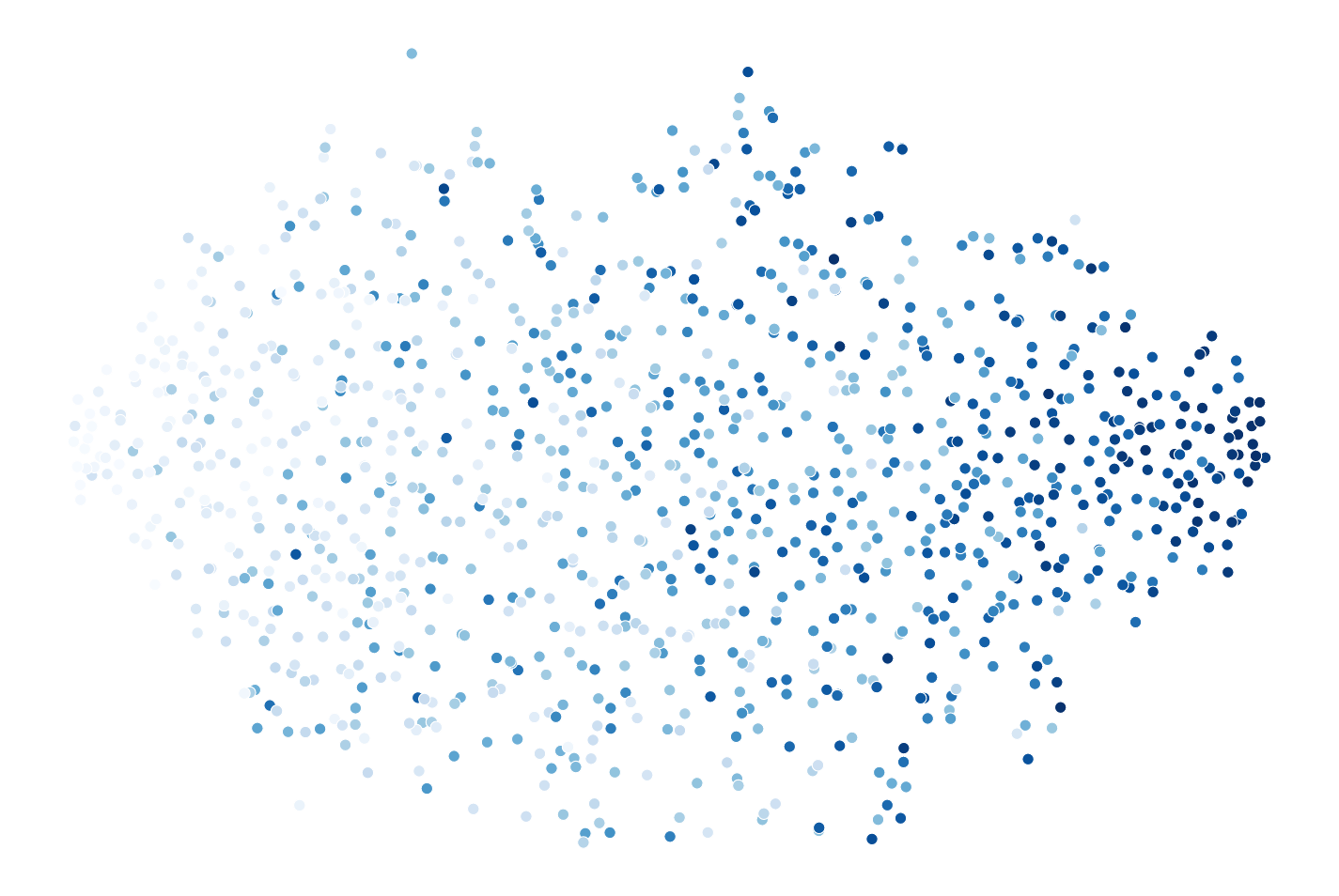}\\
    (c) KaNCD
\end{minipage}
\begin{minipage}{0.44\linewidth}\centering
    \includegraphics[width=\textwidth]{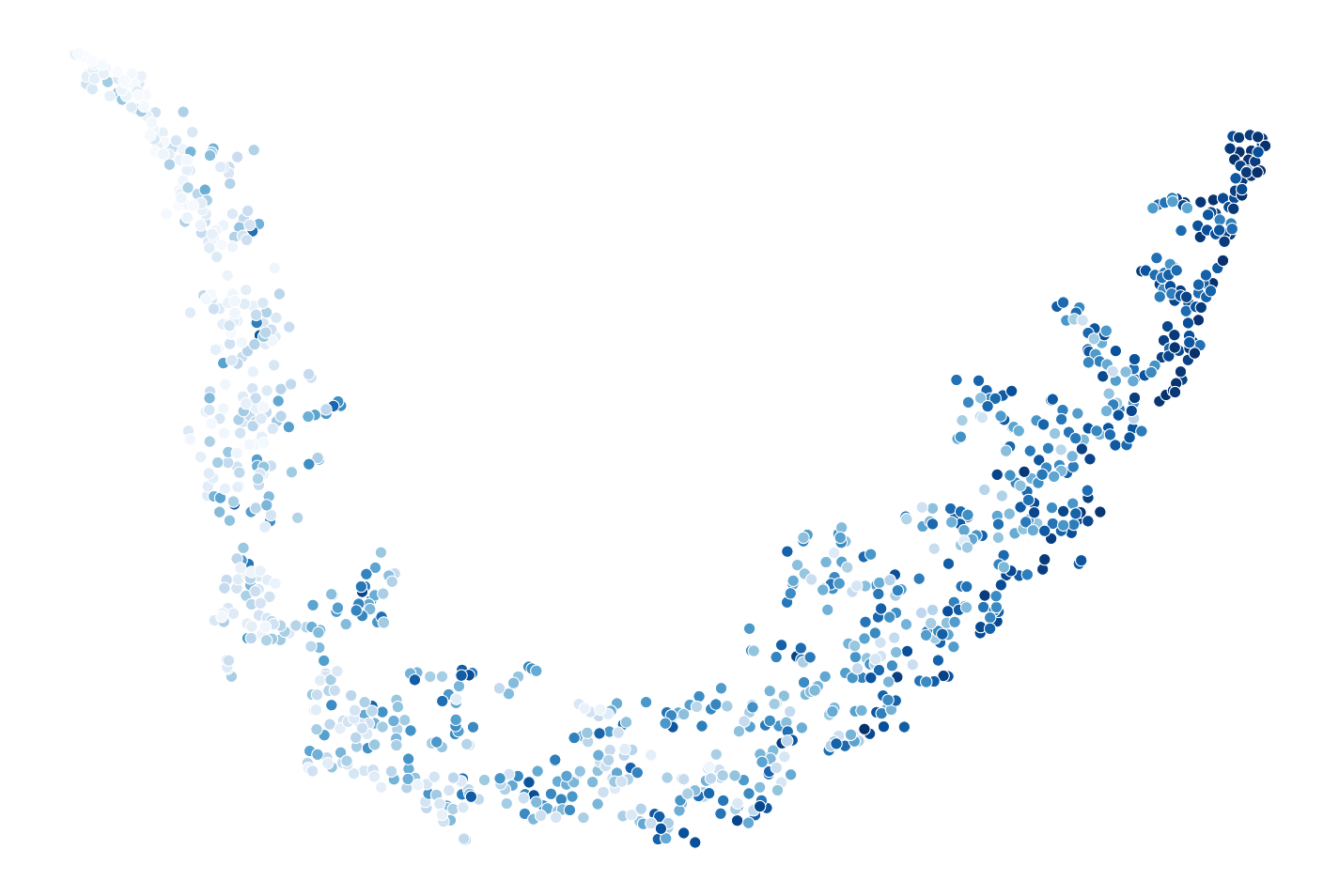}
    (d) OR-KaNCD
\end{minipage}
\\
\caption{Visualizations of the learned Mas on the EdNet-1 dataset.}
\label{fig:tsne}
\end{figure}

\textbf{The Distrubution of Students' Mas.}\label{sec:stu} Indeed, students can naturally be grouped into categories based on their performance, such as those with low and high correct rates. This classification reflects intrinsic differences in their Mas. We employ t-SNE~\cite{tsne}, a renowned dimensionality reduction method, to map the $\textbf{Mas}$ onto a two-dimensional plane. By shading the scatter plot according to the corresponding correct rates, with deeper shades of blue indicating higher correct rates, we achieve a visual representation of the students' Mas distribution. From Figure~\ref{fig:tsne}, it is clear that OR-NCDM and OR-KaNCD clusters all students $S$ with high accuracy rates more cohesively than NCDM and KaNCD.
\begin{table}[!tbp]
  \centering
  \caption{Performance in computerized adaptive testing.}
    \resizebox{0.90\linewidth}{!}{ 
    \begin{tabular}{c|l|ccc}
    \toprule
    \multicolumn{2}{c}{Dataset} & \multicolumn{3}{c}{Math2} \\
    \midrule
    \multicolumn{1}{r}{} &       & \multicolumn{3}{c}{Metric@Step} \\
    \midrule
    \multicolumn{1}{l|}{Strategy} & CDM   & AUC/ACC@5 & AUC/ACC@10 & AUC/ACC@15 \\
    \midrule
    \multirow{4}[2]{*}{Random} & IRT   & 70.80/64.35 & 74.51/66.76 & 77.63/69.01 \\
          & OR-IRT & \textbf{76.58/68.86} & \textbf{78.73/70.11} & \textbf{81.12/72.47} \\
          & NCDM  & 70.16/63.92 & 73.89/66.36 & 77.39/69.17 \\
          & OR-NCDM & \textbf{77.10/69.03} & \textbf{79.72/71.06} & \textbf{81.78/73.25} \\
    \midrule
    \multirow{4}[2]{*}{MAAT} & IRT   & 70.53/63.67 & 74.49/66.78 & 77.80/69.15 \\
          & OR-IRT & \textbf{77.89/70.26} & \textbf{79.74/72.02} & \textbf{81.59/71.81} \\
          & NCDM  & 71.35/64.85 & 74.50/66.90 & 77.25/69.26 \\
          & OR-NCDM & \textbf{78.76/69.38} & \textbf{80.53/72.00} & \textbf{81.90/73.23} \\
    \midrule
    \multirow{4}[2]{*}{BECAT} & IRT   & 70.66/64.06 & 74.36/65.42 & 77.68/68.72 \\
          & OR-IRT & \textbf{76.11/68.34} & \textbf{78.75/70.22} & \textbf{81.71/73.09} \\
          & NCDM  & 70.23/65.74 & 74.93/68.93 & 77.99/69.82 \\
          & OR-NCDM & \textbf{76.48/68.16} & \textbf{79.55/70.81} & \textbf{81.62/73.13} \\
    \bottomrule
    \end{tabular}%
    }
  \label{tab:addlabel}%
\end{table}
\subsection{Validation on the Downstream Task}
\label{exp:cat}
As an upstream task in the field of intelligent education, CD is
applied in various downstream tasks. To validate the effectiveness
of ORCDF, we chose to test it in the context of computerized adaptive testing (CAT)~\cite{ncat, zhuang2023BECAT}. Specifically, we integrate
the commonly employed IRT and NCDM with our ORCDF, denoting these as OR-IRT and OR-NCDM, respectively. Our experimental
settings align with recent research~\cite{zhuang2023BECAT}, which adopts a 7:2:1 split
for students in the response logs of each dataset. Details can be found in Appendix~\ref{appd:C}. As illustrated in Table 4, OR-NCDM performs better
than OR-IRT, which validates the superiority of deep learning-based
methods in CAT which is consistent with~\cite{ncat, zhuang2023BECAT}. OR-IRT and OR-NCDM significantly outperform their original versions. This validates the effectiveness of ORCDF in downstream tasks.

\subsection{Hyperparameter Analysis}\label{exp:hyper}

\textbf{Effect of $L$}. As shown in Figure~\ref{fig:hyper_L} in Appendix~\ref{appd:D}, a larger $L$ decreases the model's training speed, while a smaller $L$ results in poor performance. The recommended values of  $L$ are 3 or 4, which can yield relatively good performance. Notably, as $L$ increases, the MND does not continually decrease, a phenomenon that seems different from what is observed in graph representation learning. We contend this could be related to the heterogeneity of the response graph and the complexity of student interactions, which we leave for future work. 

\textbf{The Effect of $p_f$}.
 As depicted in Figure~\ref{fig:flip_ratio} in Appendix~\ref{appd:D}, OR-NCDM is influenced by the flip ratio parameter. A too high flip ratio introduces more noise, deteriorating the model's performance. Typically, a $p_f$ 
 =0.15 yields better prediction performance, aligning with the established fact that everyone has a probability of guessing correctly or slipping, neither too high nor too low.
 
\textbf{The Effect of $\lambda_{\text{reg}}$}.
As illustrated in Figure~\ref{fig:reg} in Appendix~\ref{appd:D}, this parameter controls the impact of guess and slip on model training, which varies across different datasets and requires tuning. It is observable that as the number of response logs in the dataset gradually increases, the optimal parameter value decreases. We recommend setting it to $1e^{-3}$.

\textbf{The Effect of $\tau$.}
As illustrated in Figure~\ref{fig:temp} in Appendix~\ref{appd:D}, the temperature parameter $\tau$ affects the similarity between representations learned from the response graph and those from the flipped response graph. As the size of the dataset gradually increases, the better temperature value also gradually increases. Here, we recommend choosing 0.5 when the number of students is small and opting for 3.0 when there is a larger student population.

\section{Conclusion}
This paper proposes an oversmoothing-resistant cognitive diagnosis framework (ORCDF), where most existing CDMs can be integrated and thus enhanced. We, for the first time, identify the oversmoothing in CD and then address it by learning students' Mas from multiple perspectives, utilizing the proposed response graph and response-aware graph convolution network. Besides, we reformulate the guess and slip problem as noise edges in the response graph and deign a loss function to alleviate the problem. As long as the oversmoothing is addressed in CD, it greatly helps provide distinctive and personalized diagnostic results for students and teachers. However, ORCDF, while effective, is still not sufficiently interpretable enough in the field of intelligent education. More interpretable methods are expected to be developed to  mitigate the oversmoothing issue explicably in cognitive diagnosis.

\begin{acks}
We would like to thank the anonymous reviewers for their constructive comments. We also would like to thank Xinyue Ma for the reliable help. The algorithms and datasets in the paper do not involve any ethical issue. This work is supported by the National Natural Science Foundation of China (No. 62106076), National Social Science Fund of China (No. BEA230071), and Science and Technology Commission of Shanghai Municipality Grant (No. 22511105901).
\end{acks}
\newpage
\bibliographystyle{ACM-Reference-Format}
\balance
\bibliography{reference}
\newpage
\appendix
\section*{Appendix}
The appendix is organized as follows:

$\bullet$ Appendix~\ref{appd:A} analyzes the ORCDF's time complexity and compares it with other frameworks. 

$\bullet$ Appendix~\ref{appd:B} presents the detailed settings of compared baselines and other details about student performance perdition.

$\bullet$ Appendix~\ref{appd:C} presents the detailed settings of the downstream tasks, namely, computerized adaptive testing.

$\bullet$ Appendix~\ref{appd:D} further supplements the analysis with additional details regarding the hyperparameter analysis.

Notably, our code is available at \url{https://github.com/lswhim/ORCDF}.
\begin{table}[!h]
    \caption{Abbreviations for terms.}
    \centering
    \resizebox{0.45\linewidth}{!}{ \begin{tabular}{l|l}
        \toprule
        \textbf{Term} & \textbf{Abbreviation} \\
        \midrule
        Mastery Levels & \textbf{Mas} \\
         \midrule
        Difficulty Levels & \textbf{Diff} \\
          \midrule
        Cognitive Diagnosis & CD \\
         \midrule
        Cognitive Diagnosis Model & CDM \\
          \midrule
        Degree of Agreement & DOA \\
        \bottomrule
    \end{tabular}}
    \label{tab:abbreviations}
\end{table}

\section{Time Complexity Analysis}\label{appd:A}
In this section, we present a detailed time complexity analysis of our proposed model OR-NCDM. We compare our time complexity with that of RCD, as RCD is the only CDM based on GNN.

\textbf{Time Complexity Analysis of ORCDF.} We take OR-NCDM as an exmaple. In OR-NCDM, we construct a response graph (ResG) $\mathcal{G}$ with three node and edge types based on $\mathbf{I}$ and $\mathbf{Q}$. Given that we do not employ the non-linear activation and feature transformation usually found in GNNs, the time complexity can be straightforwardly computed as \( O(2|\mathcal{E}|Ld) \)  for RGC, where $L$ denotes the number of RGC' layers. $d$ stands for the size of the embeddings. Due to the need for computing representations through the flipped ResG, the total time complexity amounts to  \( O(4|\mathcal{E}|Ld) \).

\textbf{Time Complexity Analysis of RCD.} In RCD, it construct three relation maps. Namely, an exercise-concept graph is constructed using $\mathbf{Q}$ and a student-exercise graph is formed using $\mathbf{I}$. Given that RCD employs the graph attention network, which necessitates the computation of attention coefficients between every pair of connected nodes, its time complexity belongs to $O(2|\mathcal{E}|LZ^2)$. Herein, $Z$ represents the number of concepts ($d\ll Z$).

OR-NCDM evidently takes less time compared to RCD due to two main reasons. Firstly, due to the transformation layer reduces the embedding dimension to $d$, where $d$ is much smaller than $Z$. Secondly, by removing complex operations like linear transformations in GNN, the graph convolution of RGC's computation become much faster than the GAT used in RCD.

In the experiment, we incorporate NCDM into all frameworks and use the speed of NCDM as the baseline, set at 1.0x. As shown in the figure, our proposed ORCDF is \textbf{18 times} faster than RCD and offers better prediction performance. When the number of knowledge concepts continuously increases, RCD tends to train too slowly and runs into out-of-memory issues, especially with large sets of knowledge concepts. In contrast, ORCDF maintains good performance, as demonstrated in scenarios like XES3G5M with 832 knowledge concepts on a single NVIDIA 3090 GPU, as detailed in Table~\ref{tab:pred}.
\begin{figure}[!h]
\centering
\begin{minipage}{0.6\linewidth}\centering
    \includegraphics[width=\textwidth]{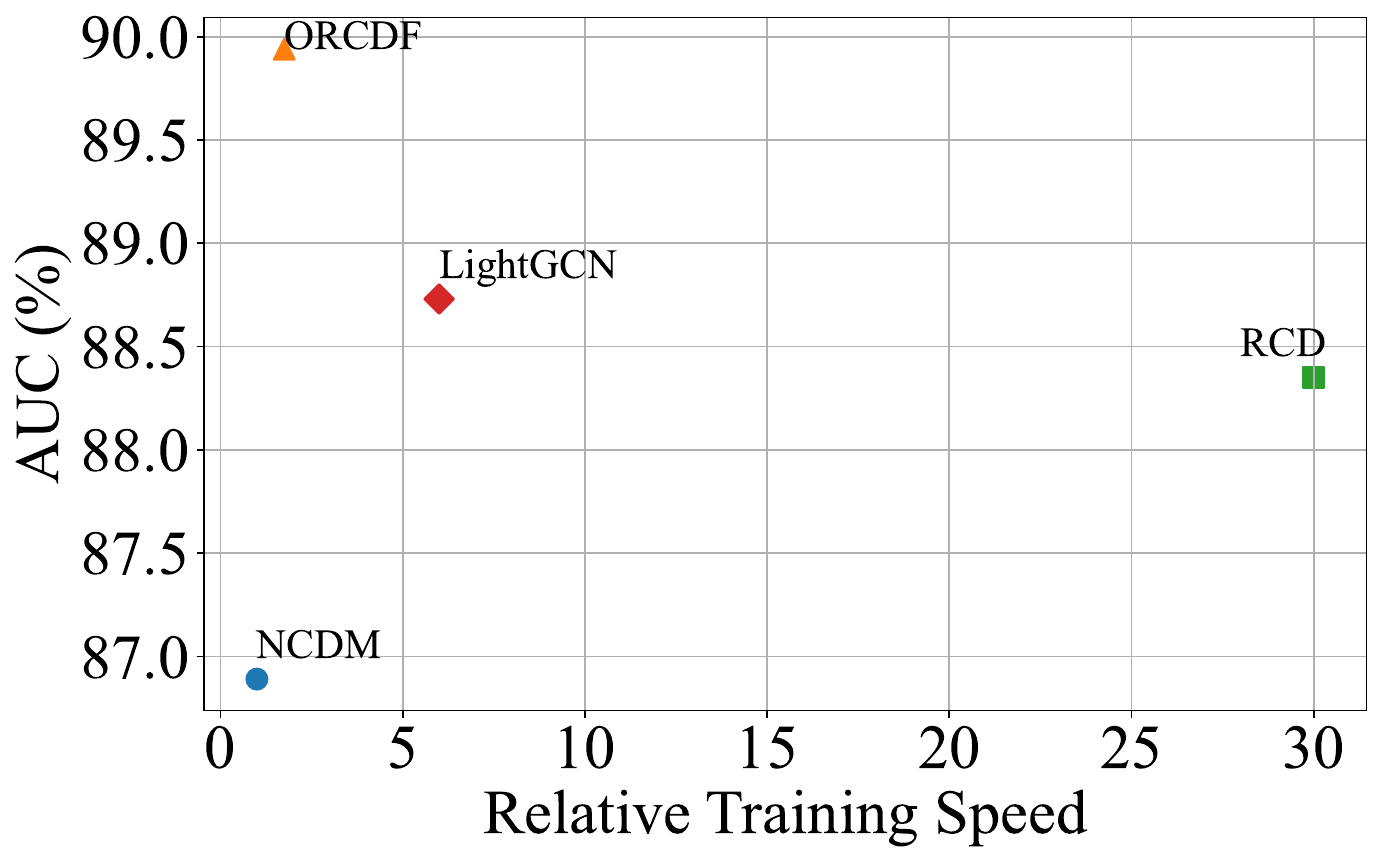}\\
\end{minipage}
\caption{Relative training speed of different related frameworks on the Assist17 dataset.}
\label{fig:TrainingTime}
\end{figure}

\section{Experimental Details}\label{appd:B}
\textbf{Interpretability Metric.}  DOA is defined as Eq.~\eqref{eq:DOA}
\begin{equation}
\label{eq:DOA}
\mathrm{DOA}_{k}=\frac{\sum_{a, b \in S} \delta\left(\mathbf{M a s}_{s_{a, c_{k}}, \mathbf{M a s}_{s_{b}, c_{k}}}\right) \frac{\sum_{j=1}^{M} \mathbf{Q}_{j k} \wedge \varphi(j, a, b) \wedge \delta\left(r_{a j}, r_{b j}\right)}{\sum_{j=1}^{M} \mathbf{Q}_{j k} \wedge \varphi(j, a, b) \wedge I\left(r_{a j} \neq r_{b j}\right)}}{Z}\,,
\end{equation}
where $Z = \sum_{a,b \in S} \delta(\mathbf{Mas}_{s_a,c_k}, \mathbf{Mas}_{s_b,c_k})$, $\mathbf{Q}_{j,k}$ indicates exercise \( e_j \)'s relevance to concept \( c_k \), \( \phi(j, a, b) \) checks if both students \( s_a \) and \( s_b \) answered \( e_j \), \( r_{a,j} \) represents the response of \( s_a \) to \( e_j \), and \( I(r_{a,j} \neq r_{b,j}) \) verifies if their responses are different, \( \delta(r_{a,j}, r_{b,j}) \) is 1 for a right response by \( s_a \) and a wrong response by \( s_b \), and 0 otherwise. 

\textbf{Implementation Details.}
This section delineates the detailed settings when comparing our method with the baselines and state-of-the-art methods in both transductive scenario and inductive scenario. All experiments are run on a Linux server with two 3.00GHz Intel Xeon Gold 6354 CPUs and one RTX3090 GPU. All the models are implemented by PyTorch. For all methods that involve using MLP as the interaction function, we adopt the commonly used two-layer tower structure with hidden dimensions of 512 and 256. Additionally, we employ the approach used in NCDM to ensure that it satisfies the monotonicity assumption.

In the following, we elaborate on some details regarding the utilization of compared methods. 
 
 $\bullet$ DINA~\cite{DINA} is a representative CDM which models the mastery pattern with discrete variables ($0$ or $1$). 

 $\bullet$ MIRT~\cite{MIRT} is a representative model of latent factor CDMs, which uses multidimensional $\bm{\theta}$ to model the latent abilities. We set the latent dimension as $16$ which is the same as~\cite{NCDM}
 
 $\bullet$ NCDM~\cite{NCDM} is a deep learning based CDM which uses MLPs to replace the traditional interaction function (i.e., logistic function). We adopt the default parameters which are reported in that paper.
 
  $\bullet$  RCD~\cite{RCD} leverages GNN to explore the relations among students,  exercises and knowledge concepts. Here, to ensure a fair comparison, we solely utilize the student-exercise-concept component of RCD, excluding the dependency on knowledge concepts.
  
 $\bullet$  KANCD~\cite{KANCD} improves NCDM by exploring the implicit association among knowledge concepts to address the problem of knowledge coverage. Here, we adopt the default parameters which are reported in that paper.
 
 $\bullet$ KSCD~\cite{kscd} also explores the implicit association among knowledge concepts and leverages a knowledge-enhanced interaction function. Due to the absence of open-source code online, we have independently replicated KSCD.

$\bullet$ LightGCN~\cite{lightgcn} is a recent classic model that employs GCN in CF. In our context, we straightforwardly consider users as students and items as exercises. We set dimension as $32$, the number of GCN layers as $3$ which is the same as OR-NCDM for a fair comparison.

 $\bullet$ HierCDF~\cite{hiercdf} is also a cognitive diagnosis framework that employs a Bayesian network, requiring a directed acyclic graph (DAG) to delineate the dependencies between knowledge concepts. It enables cognitive diagnosis models to learn mastery levels that adhere to the DAG structure, better aligning with the assumption of relationships between knowledge concepts in educational theory. We use the hyperparameters recommended in the original paper.

 The implementation of DINA, MIRT, NCDM and KANCD comes from the public repository \url{https://github.com/bigdata-ustc/EduCDM}. For RCD, we adopt the implementation from the authors in \url{https://github.com/bigdata-ustc/RCD}. 
For LightGCN, we also use the code from the authors \url{https://github.com/gusye1234/LightGCN-PyTorch}. For HierCDF, we also use the code from the authors \url{https://github.com/CSLiJT/HCD-code}.

\section{Details About Computerized Adaptive Testing.}\label{appd:C}
Computerized adaptive testing (CAT) primarily comprises CDM and item selection strategies. Its aim is to accurately determine students' mastery levels (Mas) with as few exercises as possible.
The core of CAT often lies in designing a more effective item selection strategy~\cite{Bi2020MAAT, zhuang2023BECAT}. They often opt for simple and classic CDMs like IRT or NCDM. However, in reality, these diagnostic models suffer from the oversmoothing issue and tend to underperform.

In this study, we employ a classic dataset, known as Math2, which consists of 3911 students, 16 exercises, and 16 concepts. This dataset has been widely used in various researches as referenced in studies such as~\cite{fuzzycdm, NCDM}. The objective of CAT is to accurately estimate a student's Mas using the fewest possible steps (i.e., the smallest number of exercises). However, as the true Mas cannot be obtained as ground truth, we, like previous methods, use the student performance prediction task to validate the learned Mas. For more detailed information, we recommend the readers refer to~\cite{Bi2020MAAT, zhuang2023BECAT}. Here, we utilize three commonly selected strategies which can be applied on both IRT and NCDM. These strategies can be formulated as follows.

$\bullet$ Random is a simply strategy which select exercises randomly for each student in CAT.

$\bullet$ MAAT~\cite{Bi2020MAAT} utilizes the proposed expected model change to select exercises that are likely to have a significant impact on the student's Mas.

$\bullet$ BECAT~\cite{zhuang2023BECAT} employs the concept of Coreset and utilizes expected gradient difference approximation to select exercises.
\begin{figure}[!h]
\centering
\begin{minipage}{0.40\linewidth}\centering
    \includegraphics[width=\textwidth]{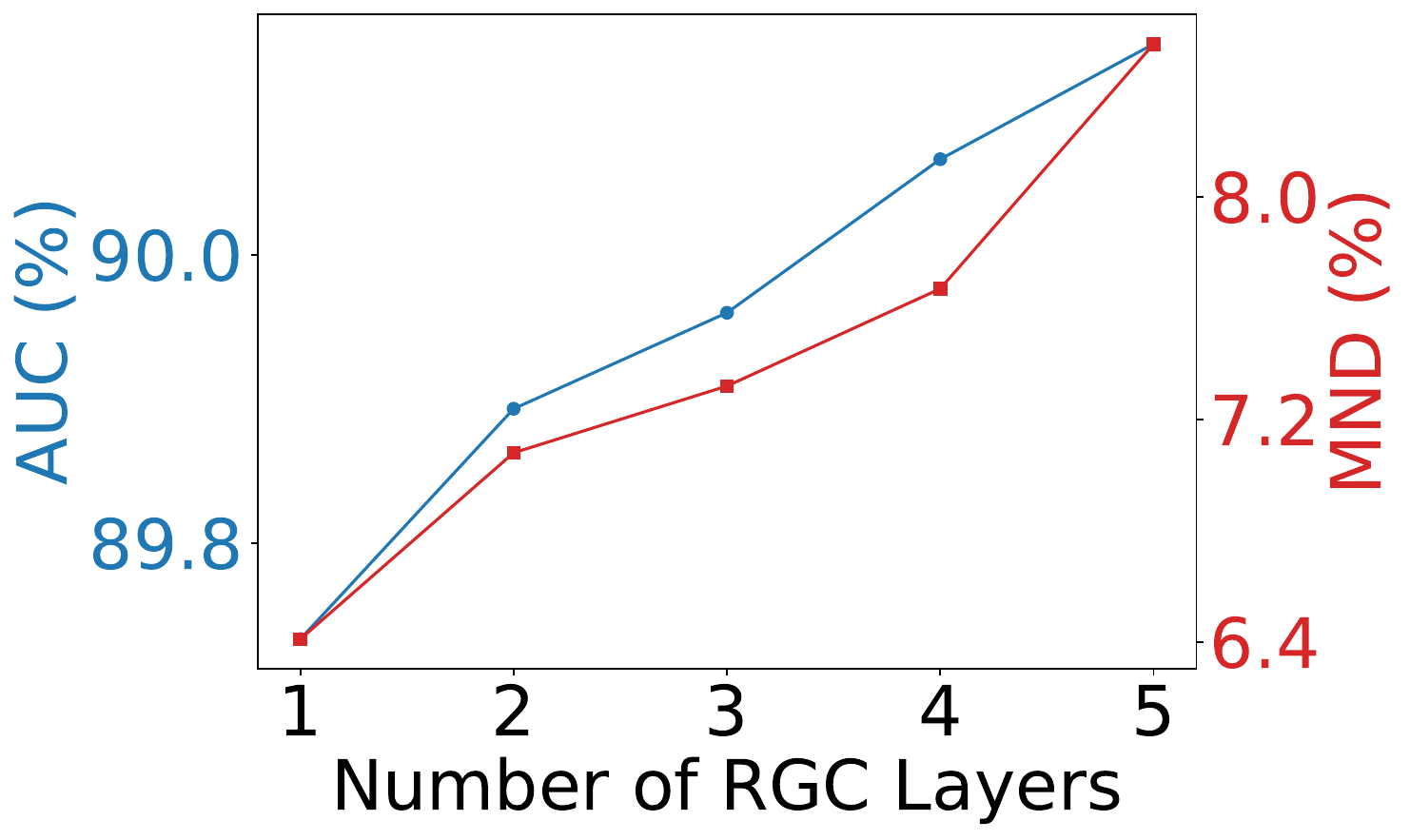}\\
\end{minipage}
\begin{minipage}{0.40\linewidth}\centering
    \includegraphics[width=\textwidth]{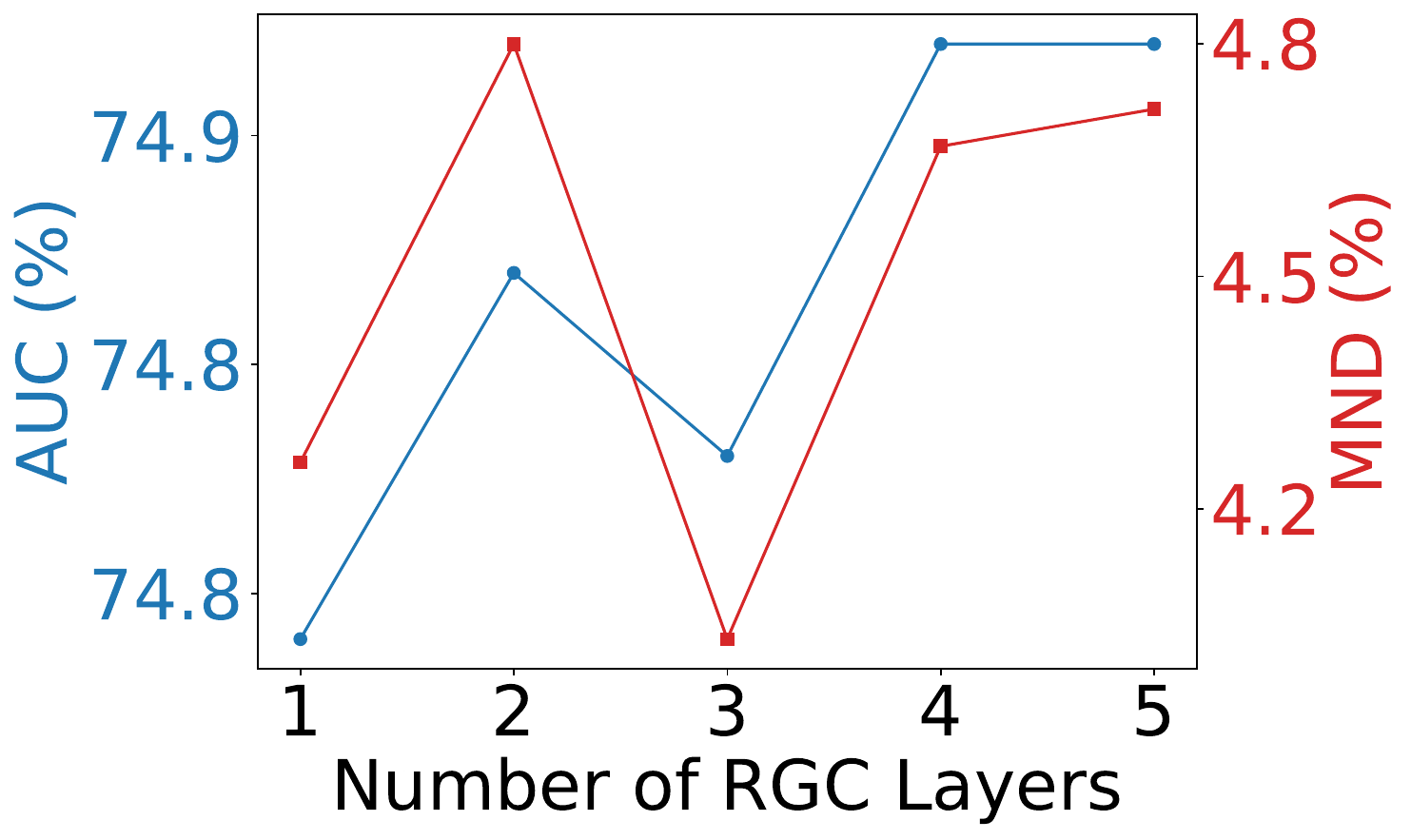}\\
\end{minipage}
\\
\begin{minipage}{0.40\linewidth}\centering
    \includegraphics[width=\textwidth]{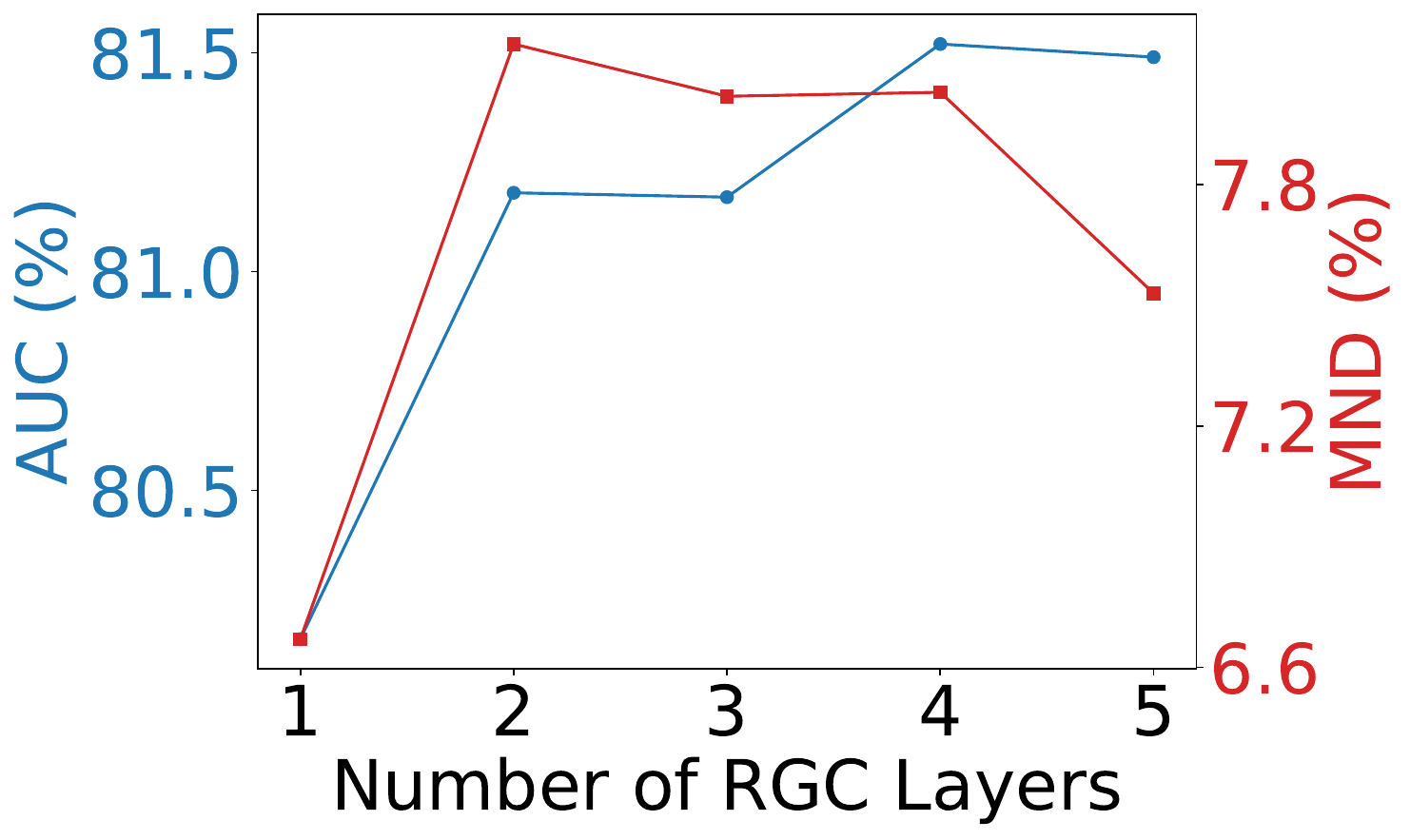}\\
\end{minipage}
\begin{minipage}{0.40\linewidth}\centering
    \includegraphics[width=\textwidth]{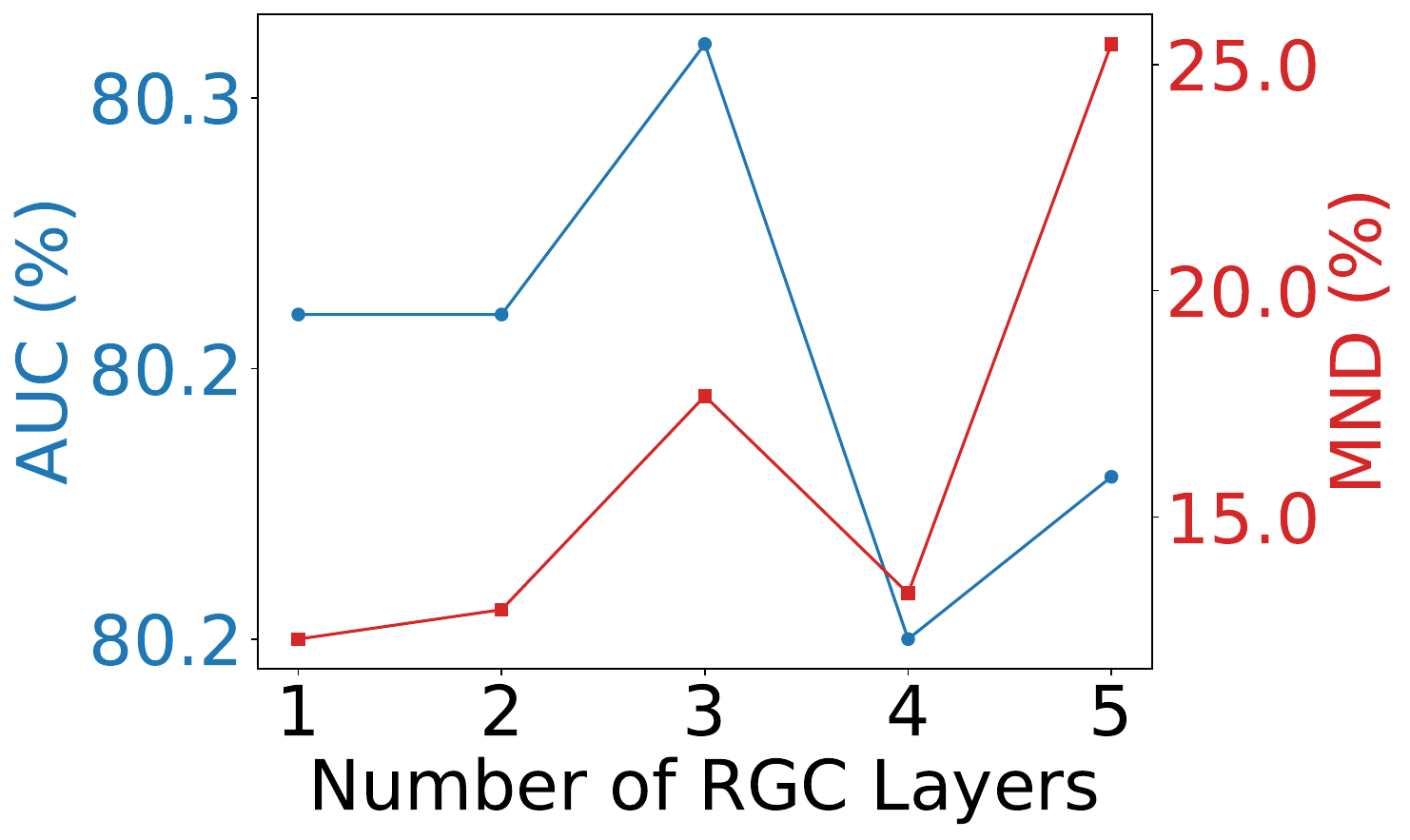}\\
\end{minipage}
\caption{Effect of $L$ on four datasets.}
\label{fig:hyper_L}
\end{figure}
\begin{figure}[!h]
\centering
\begin{minipage}{0.40\linewidth}\centering
    \includegraphics[width=\textwidth]{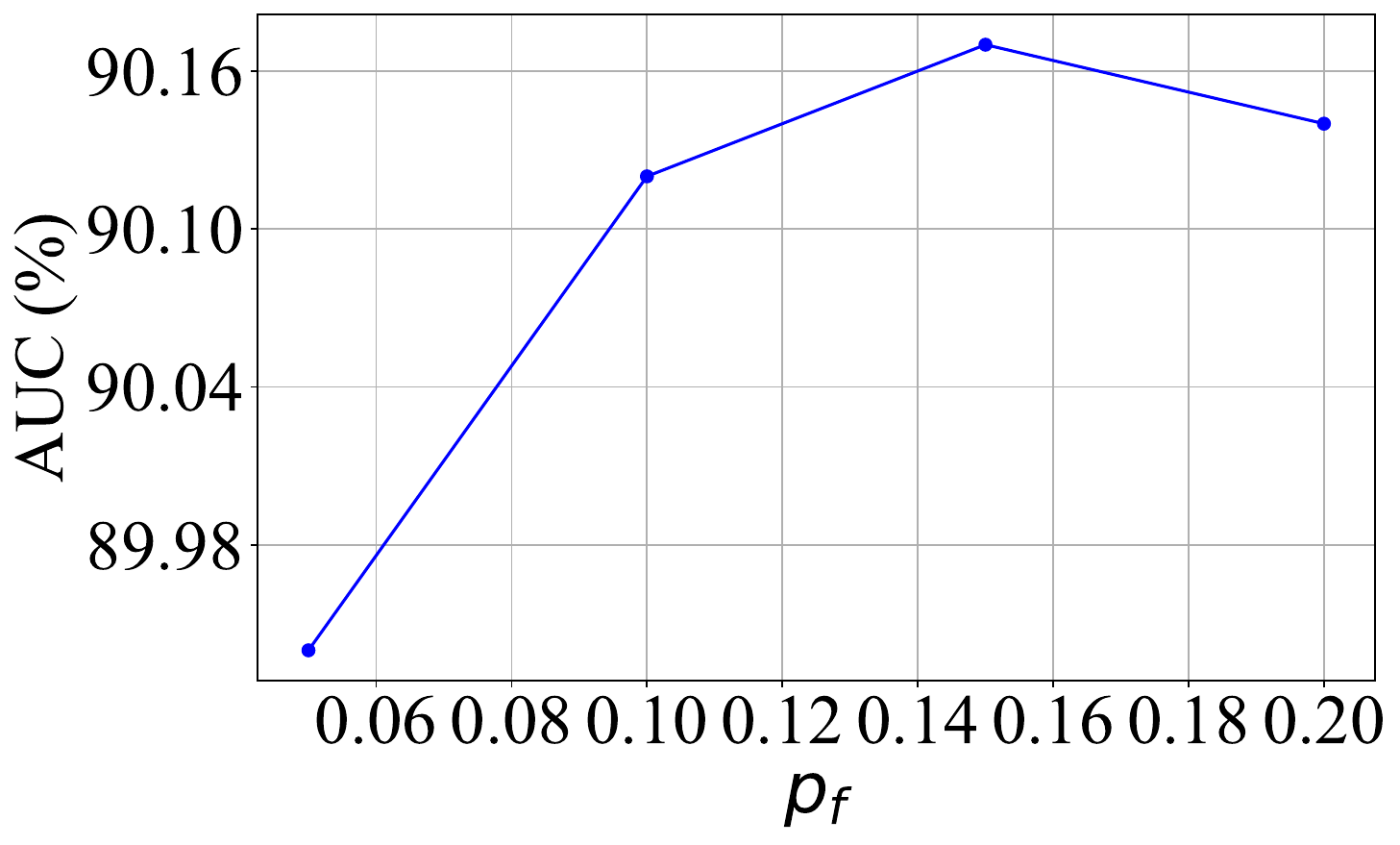}\\
\end{minipage}
\begin{minipage}{0.40\linewidth}\centering
    \includegraphics[width=\textwidth]{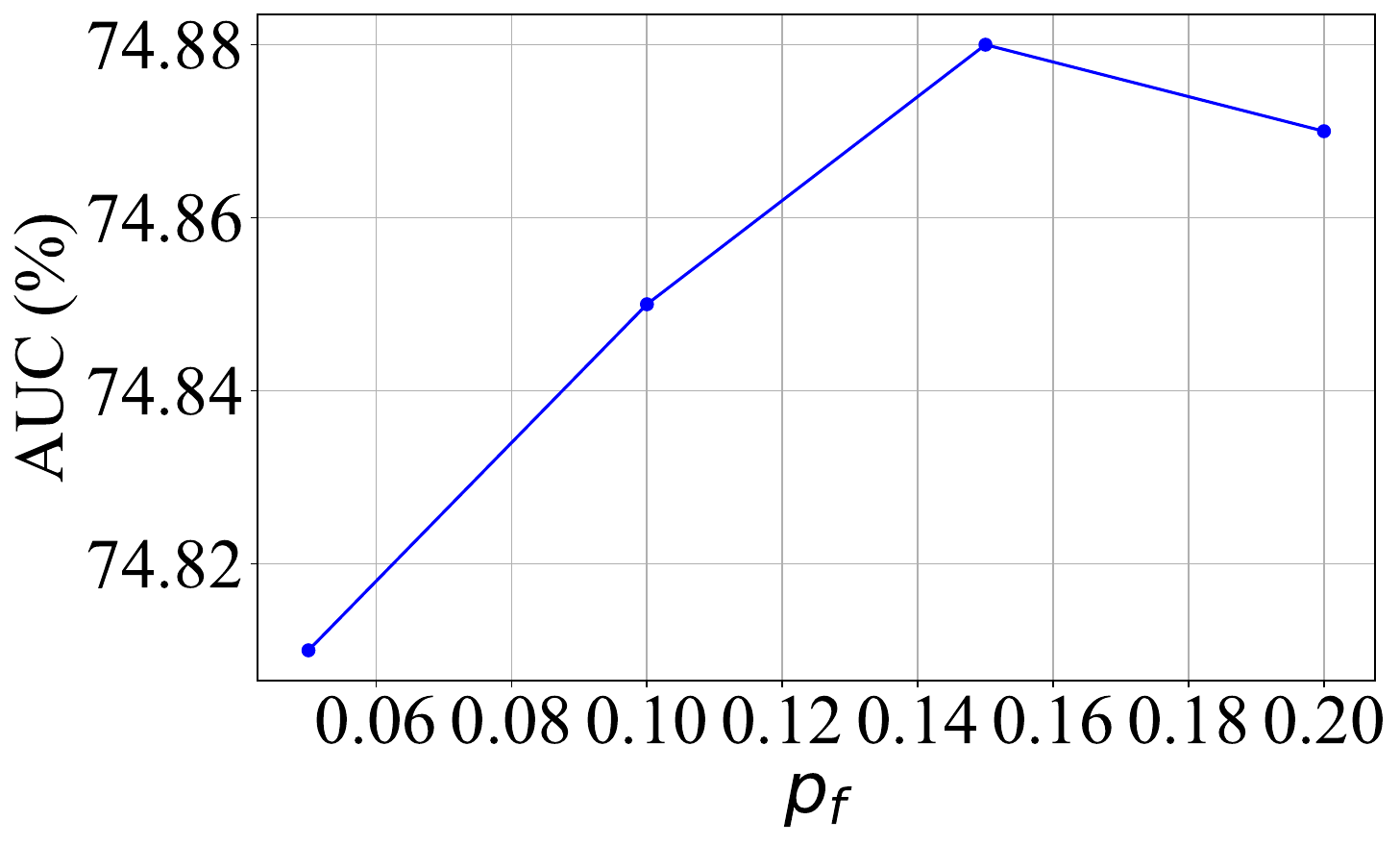}\\
\end{minipage}
\\
\begin{minipage}{0.40\linewidth}\centering
    \includegraphics[width=\textwidth]{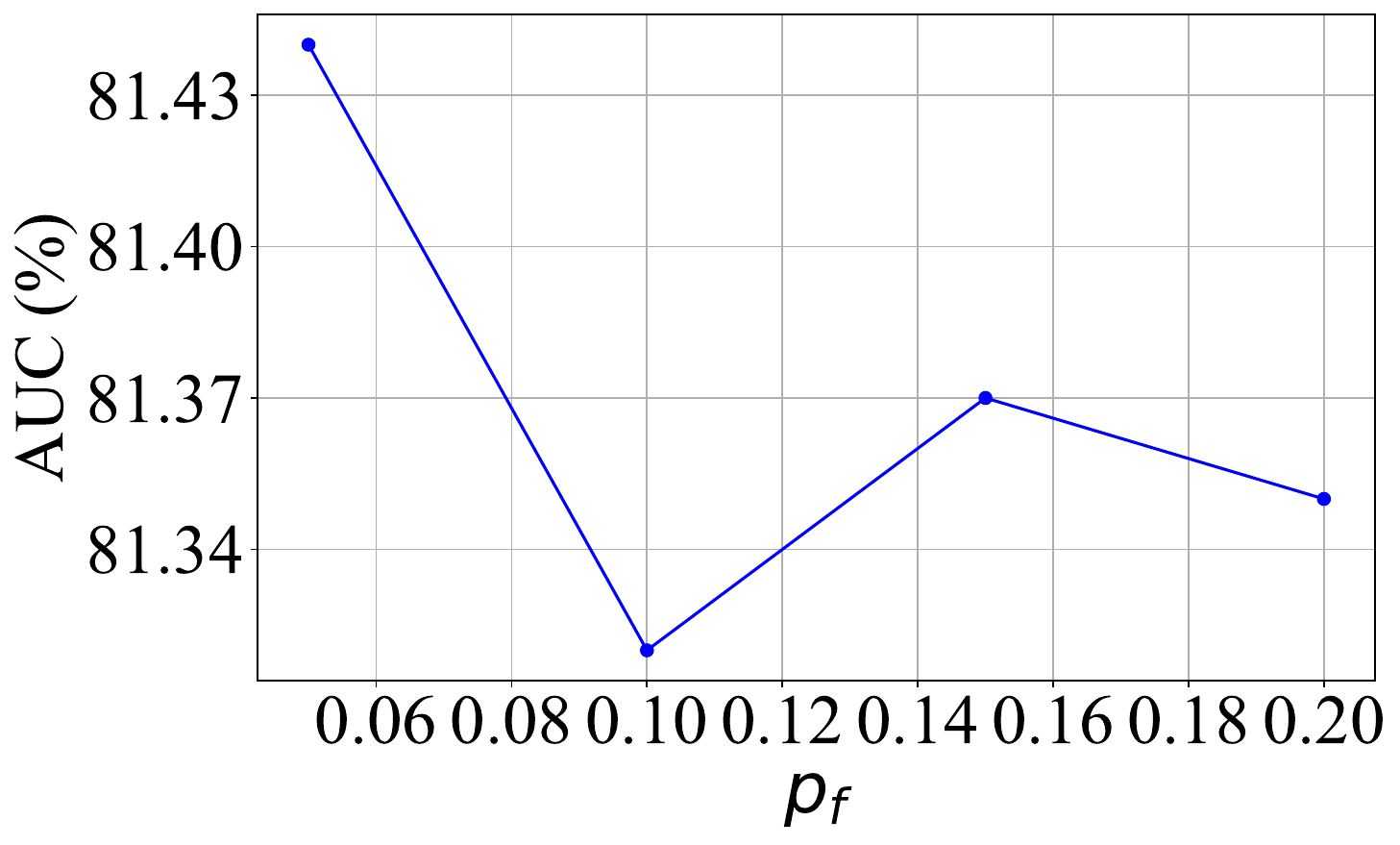}\\
\end{minipage}
\begin{minipage}{0.40\linewidth}\centering
    \includegraphics[width=\textwidth]{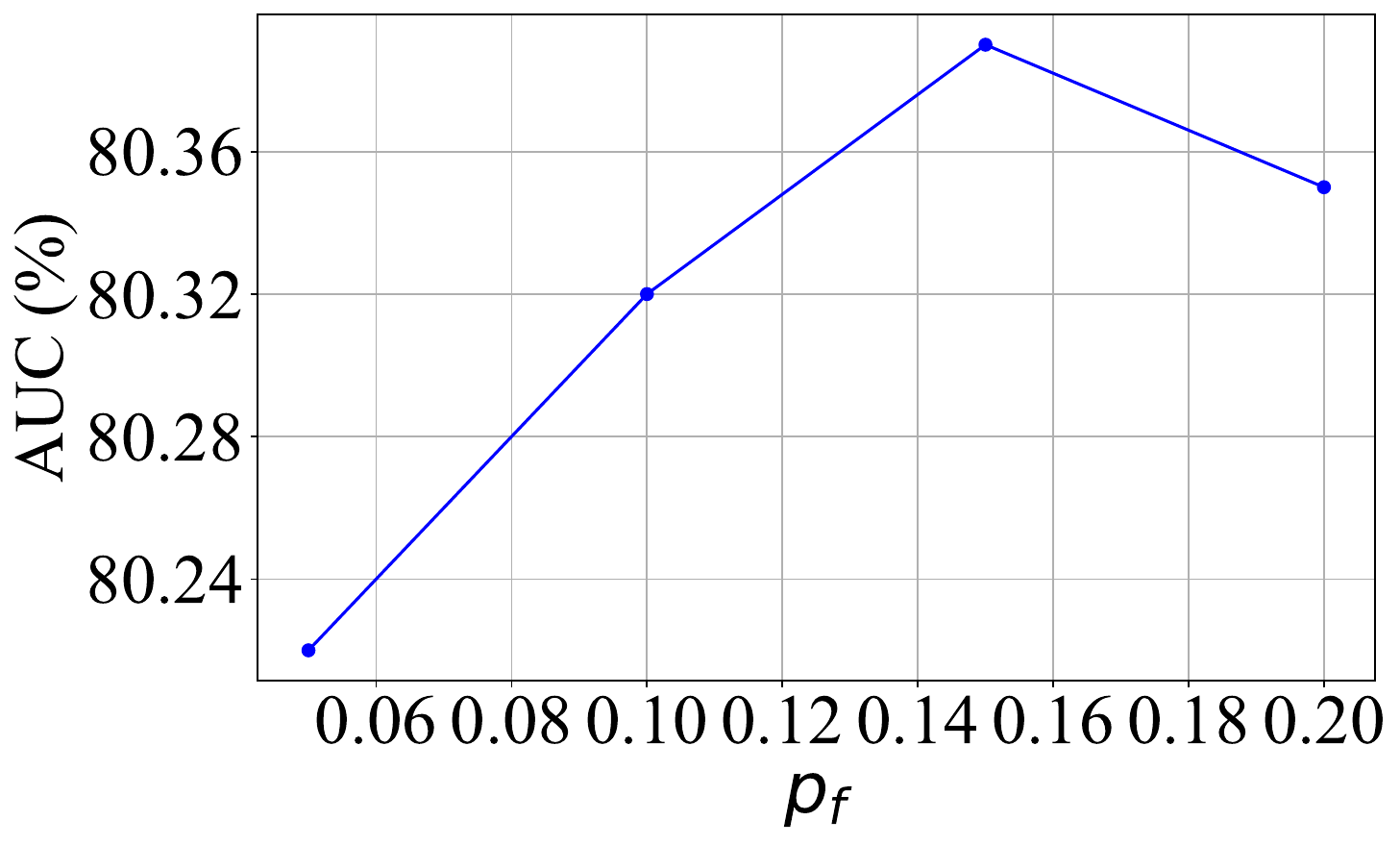}\\
\end{minipage}
\caption{Performance under different $p_f$ on four datasets.}
\label{fig:flip_ratio}
\end{figure}

\begin{figure}[!t]
\centering
\begin{minipage}{0.40\linewidth}\centering
    \includegraphics[width=\textwidth]{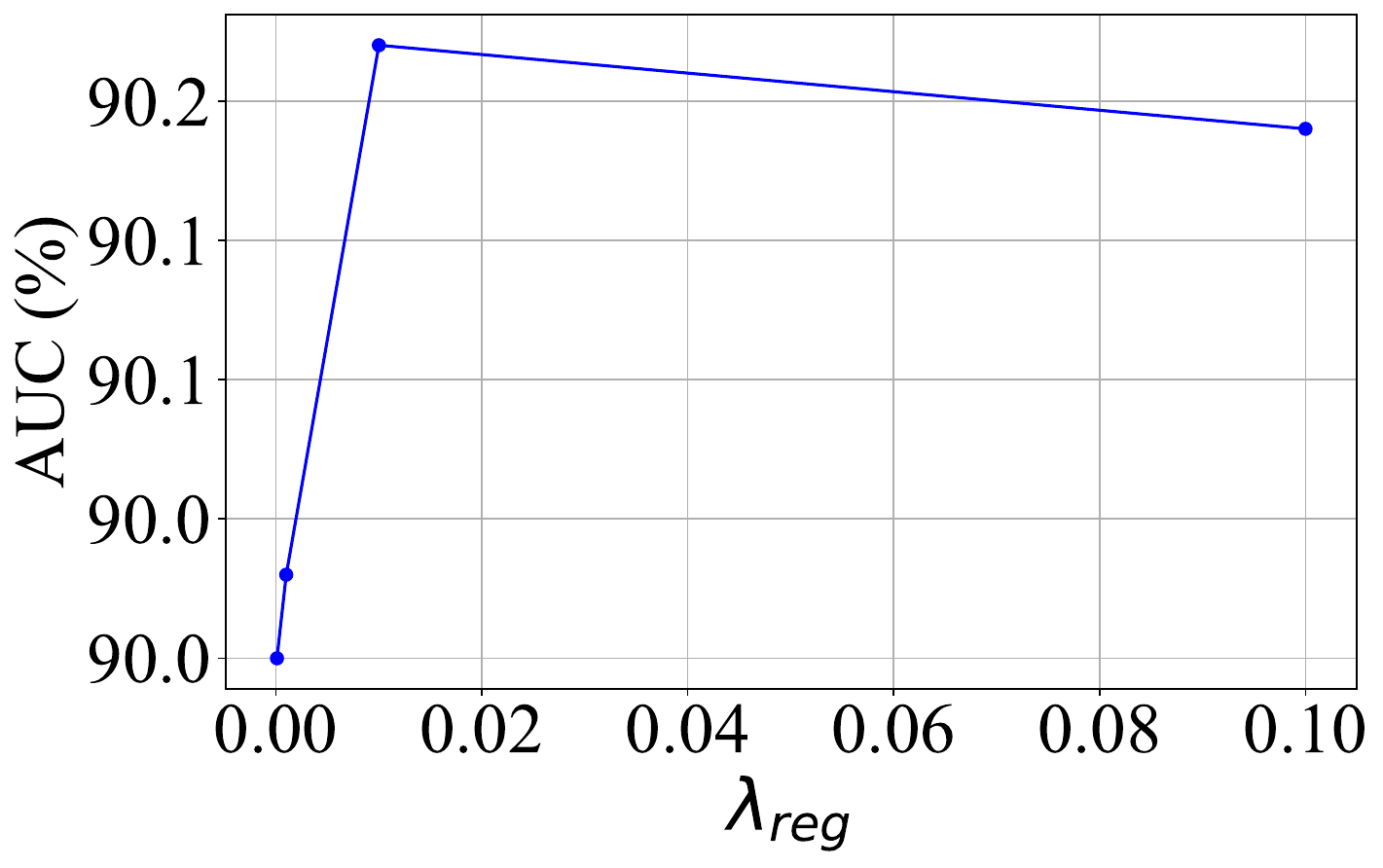}\\
\end{minipage}
\begin{minipage}{0.40\linewidth}\centering
    \includegraphics[width=\textwidth]{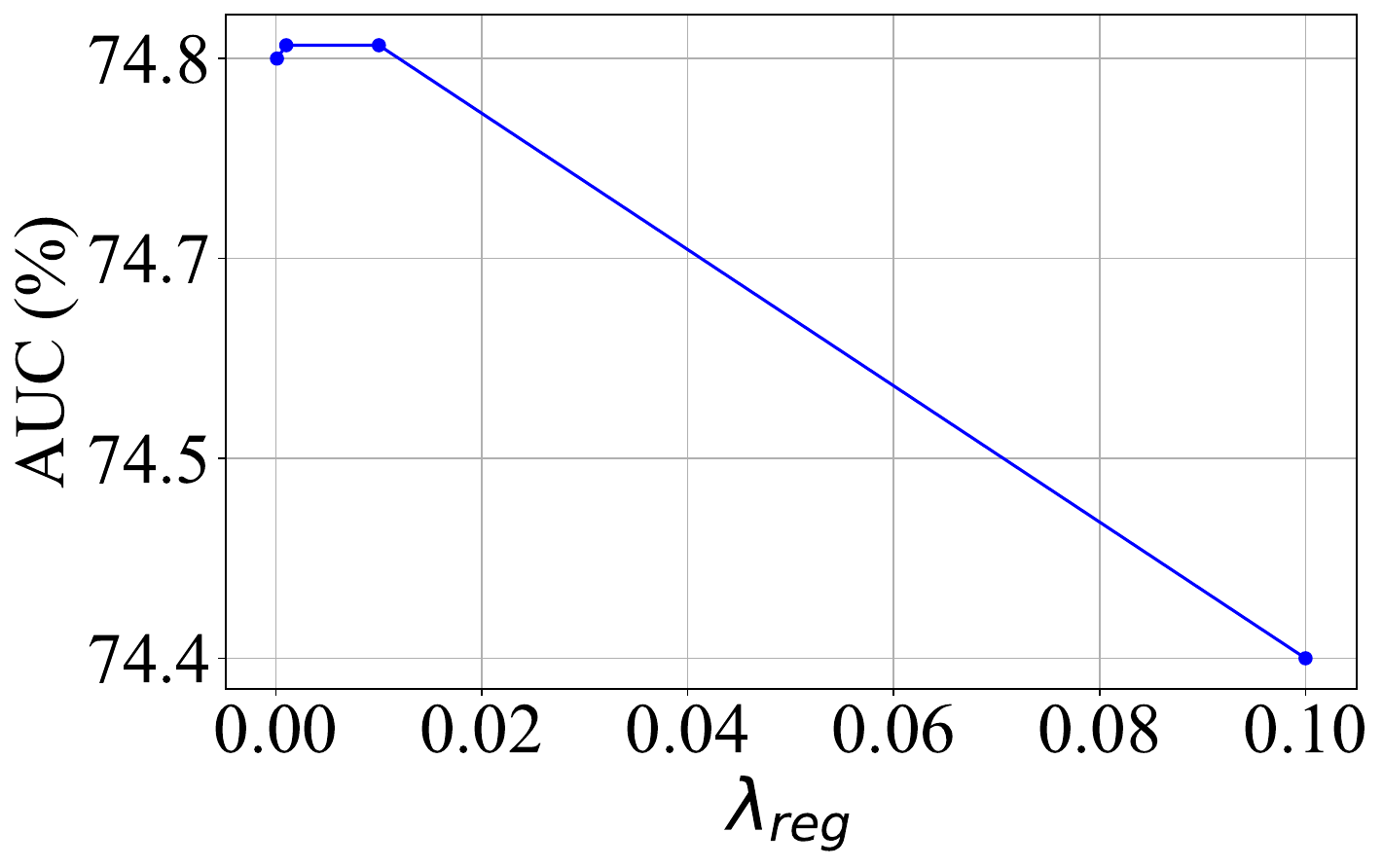}\\
\end{minipage}
\\
\begin{minipage}{0.40\linewidth}\centering
    \includegraphics[width=\textwidth]{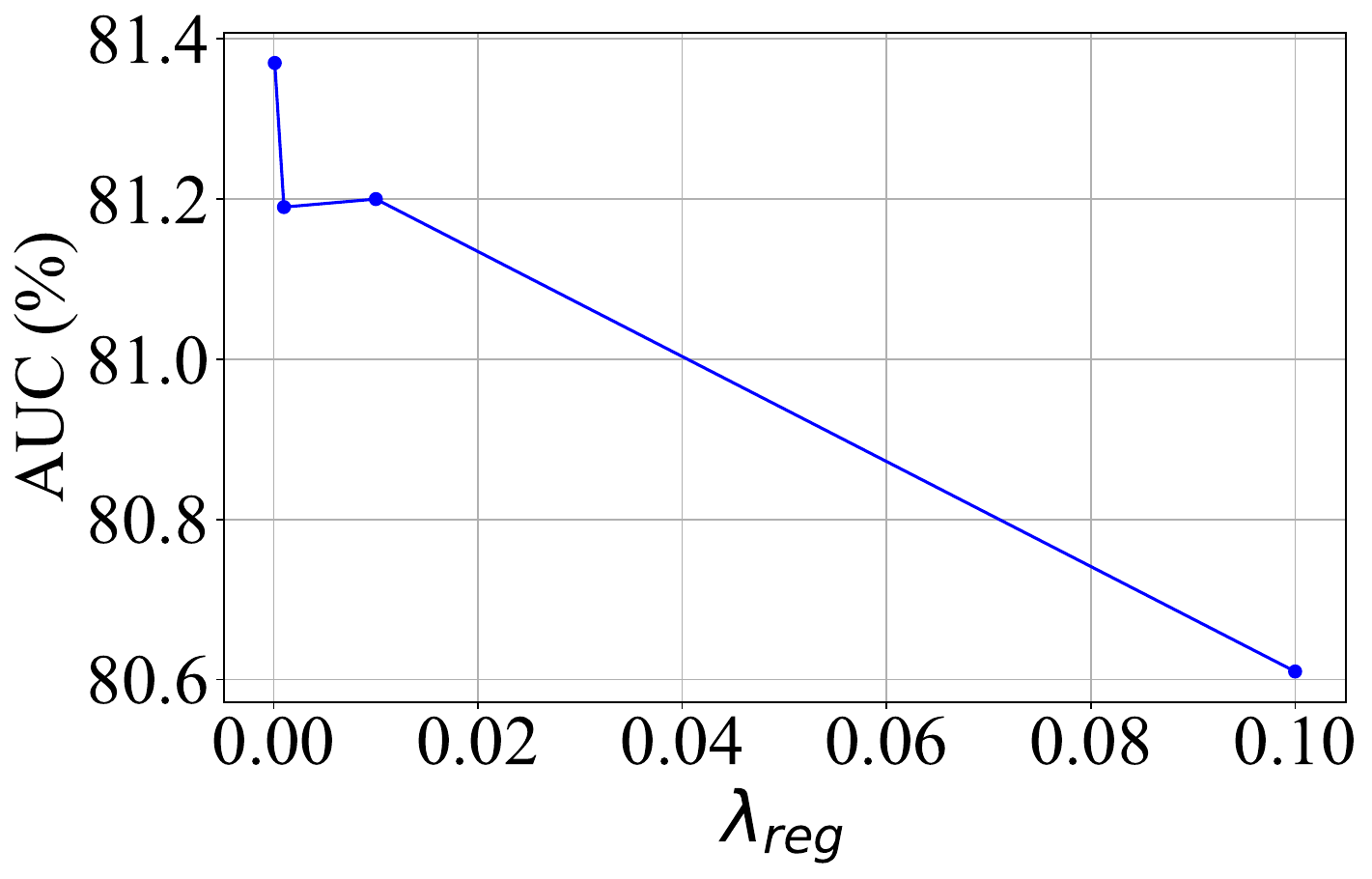}\\
\end{minipage}
\begin{minipage}{0.40\linewidth}\centering
    \includegraphics[width=\textwidth]{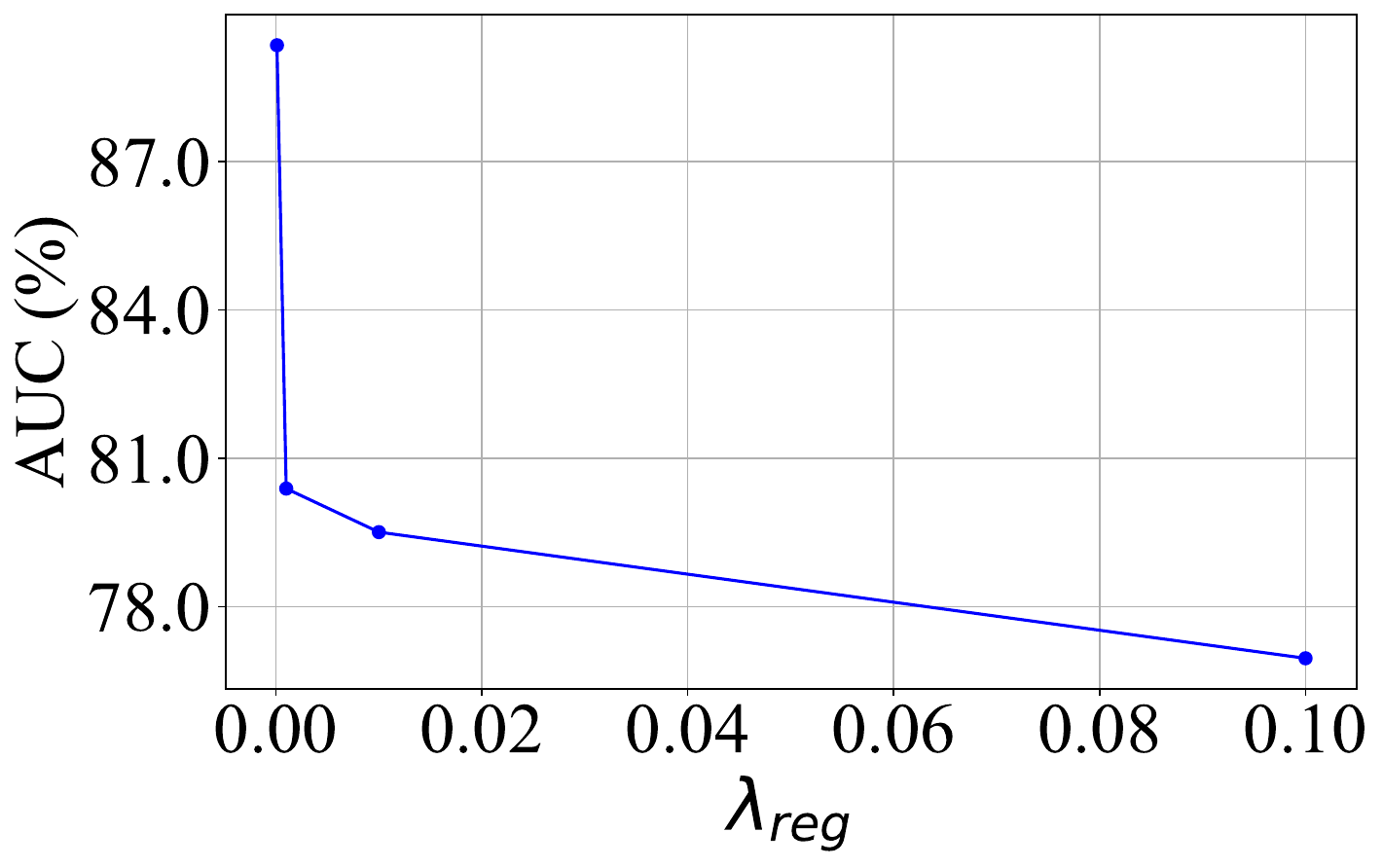}\\
\end{minipage}
\caption{Performance under different $\lambda_{\text{reg}}$ on four datasets.}
\label{fig:reg}
\end{figure}

\begin{figure}[!t]
\centering
\begin{minipage}{0.40\linewidth}\centering
    \includegraphics[width=\textwidth]{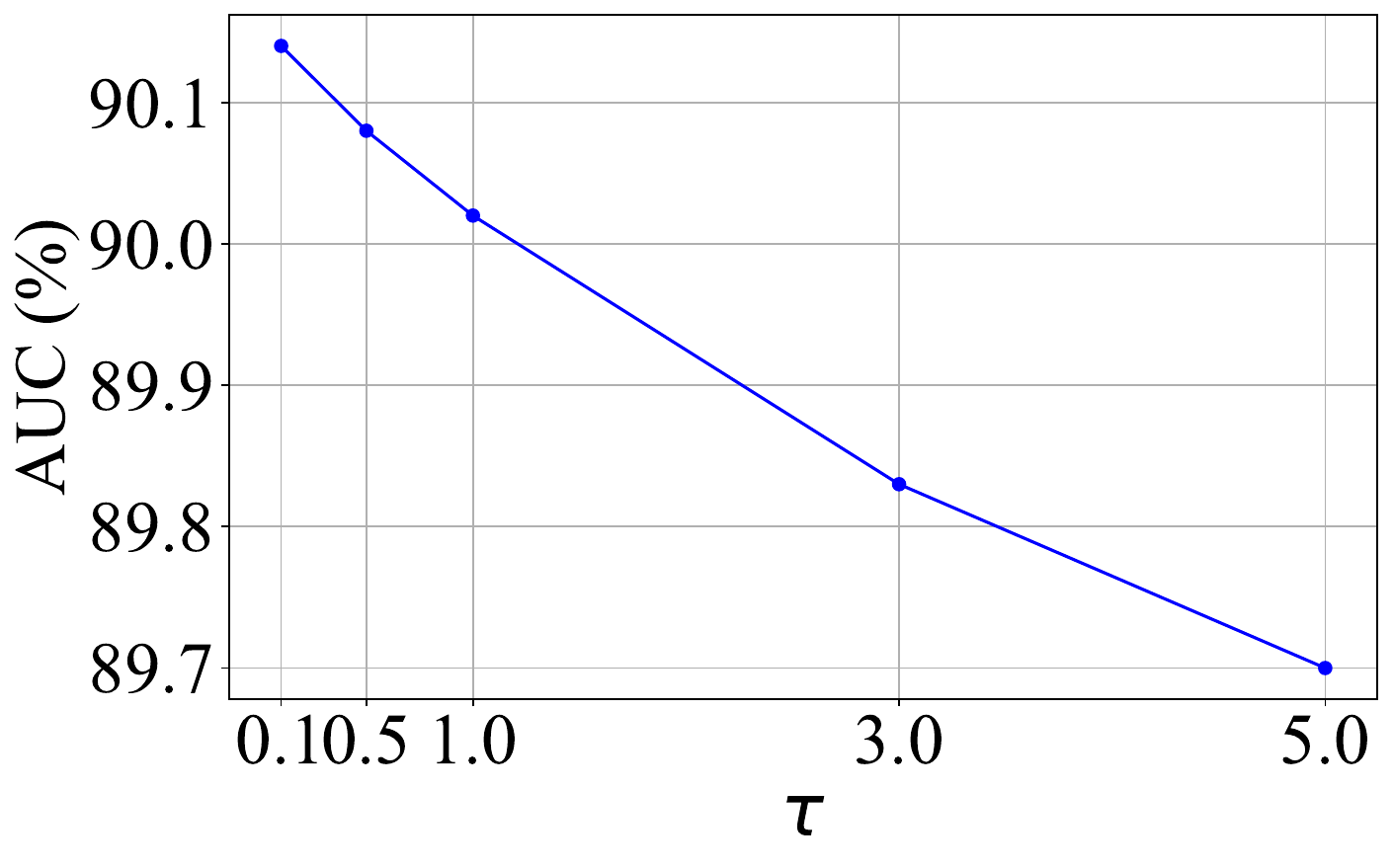}\\
\end{minipage}
\begin{minipage}{0.40\linewidth}\centering
    \includegraphics[width=\textwidth]{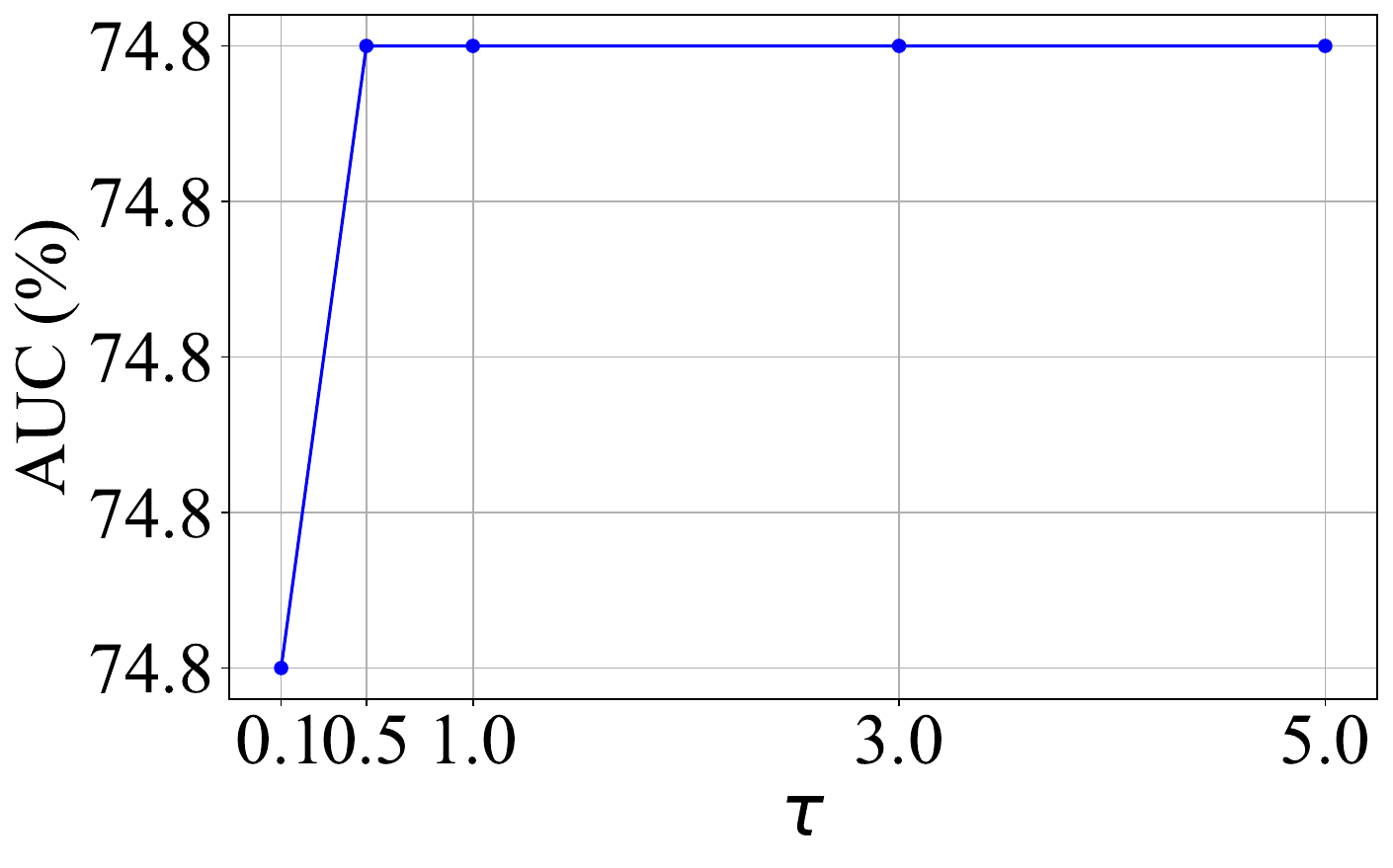}\\
\end{minipage}
\\
\begin{minipage}{0.40\linewidth}\centering
    \includegraphics[width=\textwidth]{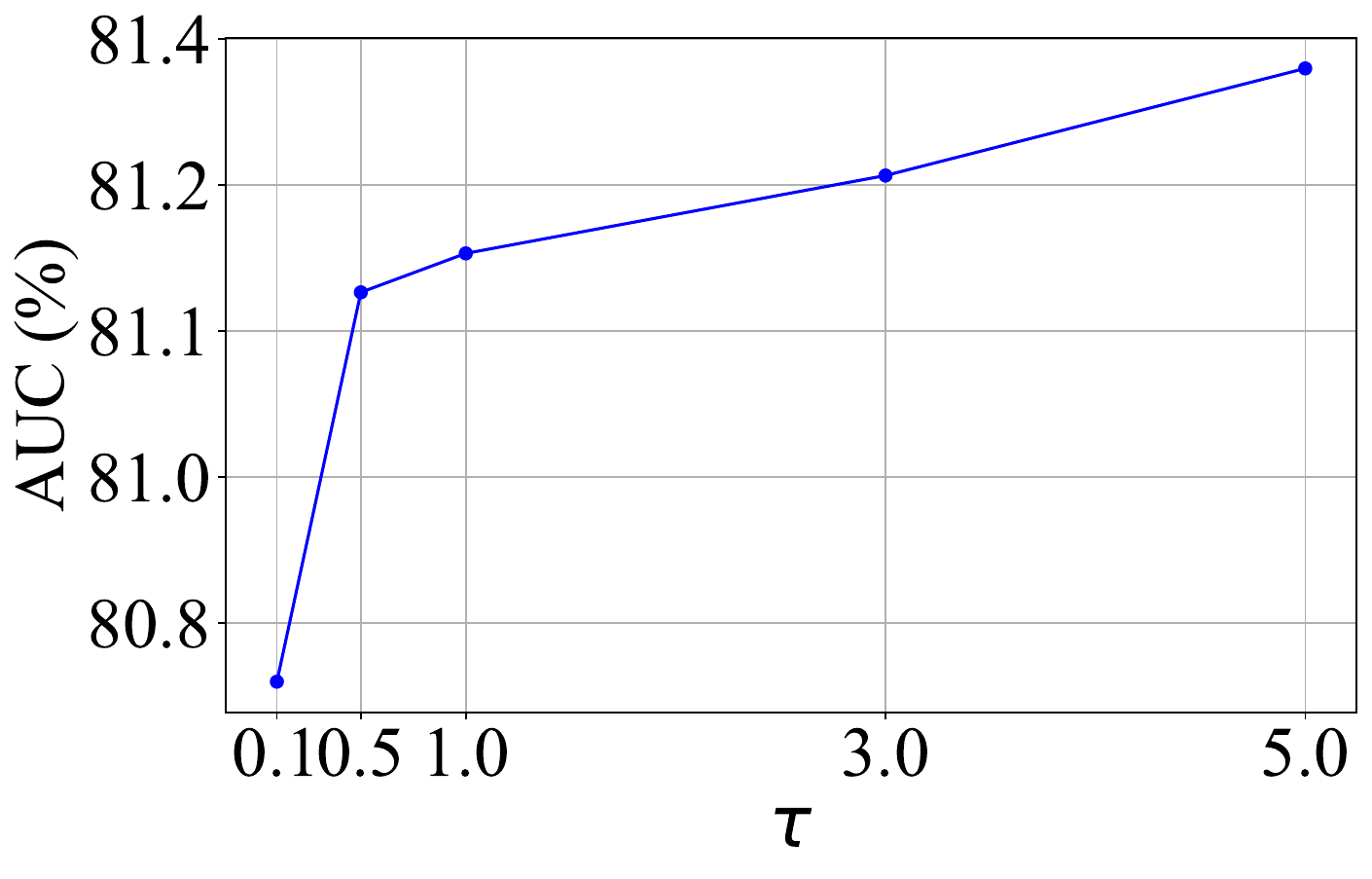}\\
\end{minipage}
\begin{minipage}{0.40\linewidth}\centering
    \includegraphics[width=\textwidth]{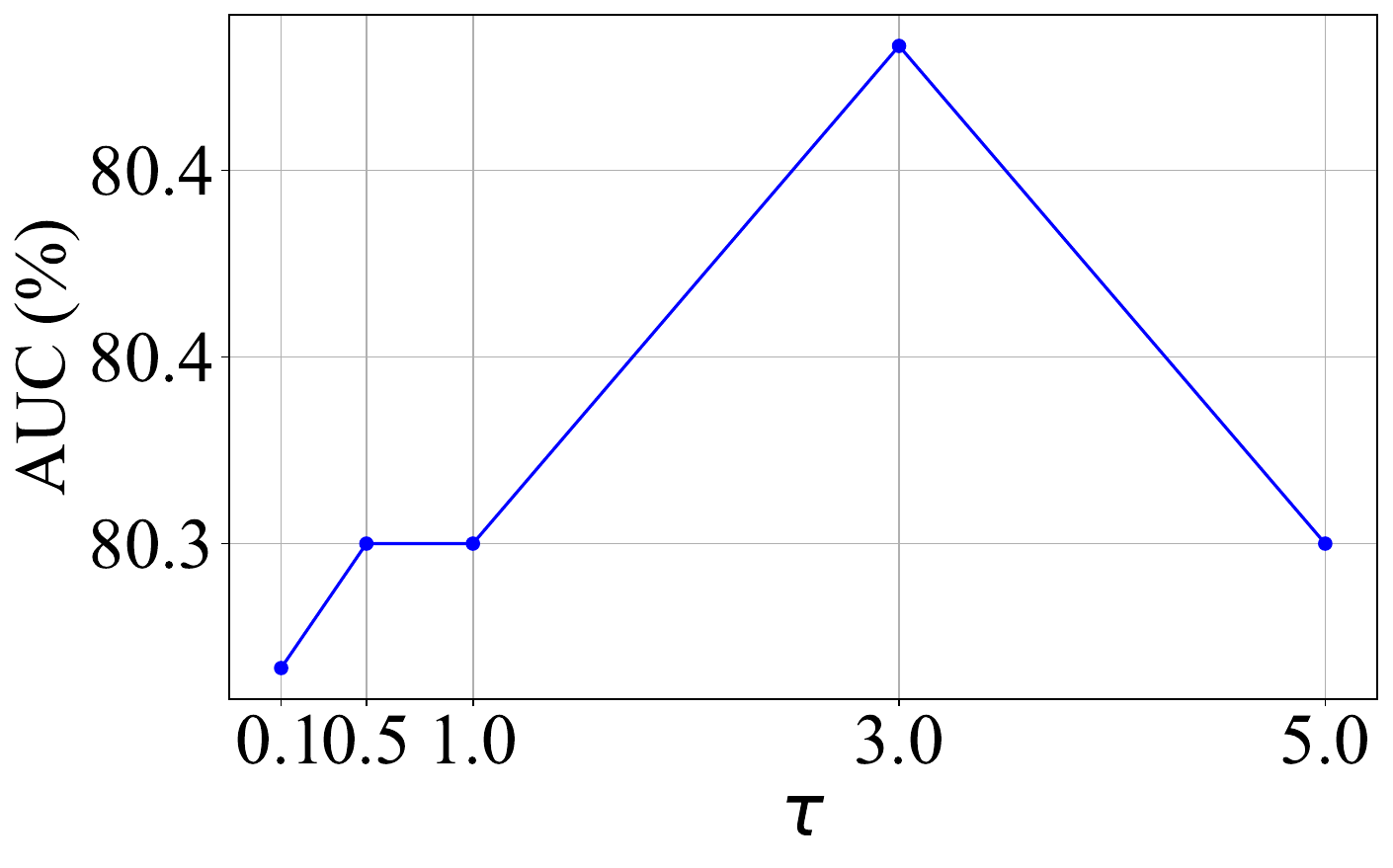}\\
\end{minipage}
\caption{Performance under different $\tau$ on four datasets.}
\label{fig:temp}
\end{figure}

\section{Details of Hyperparamter Analysis}\label{appd:D}
All figures correspond to datasets are in the order of Assist17, EdNet-1, Junyi and XESG35M. In all analyses regarding hyperparameters, we use OR-NCDM as an example.

\end{document}